%% file: kmc_dbi_i.tex
\pdfoutput=1
\documentclass[a4paper,11pt]{article}
\usepackage{jcappub} 
\usepackage{color}
\usepackage{xcolor}
\usepackage{amsmath}
\usepackage{amsfonts}
\usepackage{graphicx}
\usepackage{amssymb}
\usepackage{epstopdf}
\usepackage{url}
\usepackage{bm}
\usepackage{hhline}
\usepackage{subfigure}
\usepackage{multirow}
\usepackage{epsfig}
\usepackage{varioref}
\usepackage{amssymb}
\usepackage{cases}
\usepackage{multirow}
\usepackage{mathrsfs}
\usepackage{calrsfs}
\usepackage{mathtools}
\usepackage{amsthm}
\usepackage{hyperref}
\def\sect{\section}
\def\subs{\subsection}
\def\ssubs{\subsubsection}

\def\be{\begin{equation}} \def\ee{\end{equation}}
\def\ba{\begin{eqnarray}} \def\ea{\end{eqnarray}}
\def\bt{\begin{tabular}} \def\et{\end{tabular}}
\def\btb{\begin{table}} \def\etb{\end{table}}
\def\bay{\begin{array}} \def\eay{\end{array}}
\def\bfg{\begin{figure}} \def\efg{\end{figure}}
\def\bc{\begin{center}} \def\ec{\end{center}}
\def\bi{\begin{itemize}} \def\ei{\end{itemize}}
\def\bl{\begin{list}} \def\el{\end{list}}
\def\bn{\begin{enumerate}} \def\en{\end{enumerate}}
  
 \def\raw{\rightarrow} 
\def\lrb{\left(} \def\rrb{\right)}
\def\lsb{\left[} \def\rsb{\right]}

\def\ul{\underline}

\def\td{\tilde}
\def\spc{\,,\,} 
\newcommand{\defeq}{\vcentcolon=}

\def\mcl{\mathcal}
\def\mbf{\mathbf}

\def\tX{\td X} 
\def\tV{\td V} \def\tp{\td p}

\def\ls{\lambda_{\rm s}}
\def\cs{c_{{\rm s}}} 
\def\css{c_{{\rm s}}^{2}} 
\def\L{\mathcal{L}} 
\def\Vp{V^{\prime}}   
\def\tV{\tilde{V}}        
\def\Wp{W^{\prime}} 

\def\r{\rho}  
\def\up{u^{\prime}} 
 
\def\Wf{W_{,\varphi}} \def\Wff{W_{,\varphi\varphi}}
    
\def\Mpl{M_{\rm Pl}^{2}} 
\def\Ps{\mathcal{P}_{\zeta}} 
\def\ns{n_{{\rm s}}-1} 
\def\dN{\delta{N}} 
\title{Observational Constraints on Tachyon and DBI inflation}
\author{Sheng Li$^1$}
\author{and Andrew R. Liddle$^{1,2}$} 
\affiliation{$^1$Astronomy Centre, University of Sussex, Brighton BN1 9QH,
United Kingdom}
\affiliation{$^2$Institute for Astronomy, University of Edinburgh, Royal Observatory, Edinburgh EH9~3HJ, United Kingdom}

\emailAdd{sl277@sussex.ac.uk}
\emailAdd{arl@roe.ac.uk}
\date{\today}

\abstract{
We present a systematic method for evaluation of perturbation observables in non-canonical single-field inflation models within the slow-roll approximation, which allied with field redefinitions enables predictions to be established for a wide range of models. We use this to investigate various non-canonical inflation models, including Tachyon inflation and DBI inflation. The Lambert $\mcl W$ function will be used extensively in our method for the evaluation of observables. In the Tachyon case, in the slow-roll approximation the model can be approximated by a canonical field with a redefined potential, which yields predictions in better agreement with observations than the canonical equivalents. For  DBI inflation models we consider contributions from both the scalar potential and the warp geometry. In the case of a quartic potential, we find a formula for the observables under both non-relativistic (sound speed $\css \sim 1$) and relativistic behaviour ($\css \ll 1$) of the scalar DBI inflaton. For a quadratic potential we find two branches in the non-relativistic $\css \sim 1$ case, determined by the competition of model parameters, while for the relativistic case $\css \raw 0$, we find consistency with results already in the literature. We present a comparison to the latest {\it Planck} satellite observations. Most of the non-canonical models we investigate, including the Tachyon, are better fits to data than canonical models with the same potential, but we find that DBI models in the slow-roll regime have difficulty in matching the data.
}

\begin{document}
\maketitle

\sect{Introduction}

\label{sect1}

The slow-roll expansion has proven a powerful tool for calculating perturbation observables in inflationary cosmologies, and its likely applicability is now strongly supported by the observed near scale-invariance of scalar perturbations and the increasingly strong upper limits on the tensor-to-scalar ratio. In the case of canonically-normalized single-field models this formalism, first set down in Ref.~\cite{LL92}, readily generates results that can be set against observations such as the recent data compilations provided by the {\it Planck} Collaboration
\cite{Ade:2013uln}. It results from an expansion in the Mukhanov equations \cite{Muk} that describe scalar and tensor perturbations, which can be solved using Hankel functions. The formalism can be extended to higher order  in the small slow-roll parameters \cite{Stewart:1993bc,Stewart:2001cd,Gong:2001he,Gong:2002cx,Choe:2004zg}), and various non-slow-roll approaches exist such as the $\dN$ formalism \cite{Sasaki:1995aw} and the approaches of Refs.~\cite{Huston:2011vt,Ribeiro:2012ar,Adshead:2013zfa}, which are often useful in non-gaussianity investigations \cite{Burrage:2011hd}. For the power spectra themselves, however, the simple slow-roll approach is typically valid.

There is ongoing interest in the possibility that inflation may be driven by a single field which does not possess a canonical kinetic term, usually referred to as k-inflation \cite{ArmendarizPicon:1999rj}, examples being the Tachyon and the Dirac--Born--Infeld (DBI) field, or the Galileon inflation theories as discussed in Refs.~\cite{Mizuno:2010ag,Kobayashi:2010cm,Burrage:2010cu}. This motivation is in large part theoretical, but additional observational impetus is being created due to the tightening upper limits on the tensor contribution, as even simple non-canonical models can shift predicted observables closer to scale invariance, e.g.\ Refs.~\cite{Li:2012vt,Unnikrishnan:2012zu,Unnikrishnan2}. The slow-roll formalism is readily extended to non-canonical single-field models \cite{Garriga:1999vw,Kinney,Ringeval} (see Ref.~\cite{Hu:2011vr} for recent developments). However, for specific models the calculations necessary to determine the observables may be algebraically challenging.

In this article we propose a new systematic algorithm for carrying out these calculations, through a reframing of the evolution equations in a recursive form and use of field redefinitions to simplify the Lagrangians. We deploy it for a range of non-canonical models to generate theoretical predictions that we set against contemporary observations. The new algorithm will frequently involve the Lambert $\mcl W$ function in order to evaluate desired observables such as the power spectrum and its index. The results from our algorithm give an interpretation of the numerical results, in particular the degeneracy of model parameters \cite{Li:2012vt,Devi:2011qm}. The algorithm also provides another view on reconstructing the inflation model in respect of the observational data, such as those from {\it Planck} and future data from {\it Euclid}.

We begin by introducing the framework of the proposed algorithm in Section~\ref{sect-methods}, and present the predictions for the power spectrum and its spectral index for various models. We implement the analytic methods in Section~\ref{sect-twomods}. We will discuss our results for some models in Section~\ref{sect-implement}, then model parameter estimations are studied in Section~\ref{sect-mpe} as the result of applying those techniques. We summarise in Section \ref{sect-conclusion}, including a few words on future exploration and possible applications of our algorithm.

\section{The Models}\label{sect-models}
The general Lagrangian for a single-field model with second-order field equation is an arbitrary function $p(X,\phi)$ of the scalar field $\phi$ and its kinetic energy $X \equiv \frac{1}{2} \partial_\mu \phi \partial^\mu \phi$. In addition to the general case, in this paper we will consider four specific Lagrangians within this class:
\begin{enumerate}
\item Canonical field with potential $V(\phi)$:
\begin{equation}
\label{m:canonical}
p(X,\phi) = BX - V(\phi) \,,
\end{equation}
where we add the coefficient $B$ denoting the coupling strength of the kinetic energy, though traditionally it is set to 1.
\item Non-canonical inflation (NCI) model: This features an arbitrary power on the kinetic term and was studied in Refs.~\cite{Mukhanov:2005bu,Li:2012vt,Unnikrishnan:2012zu,Unnikrishnan2}, the Lagrangian being
\begin{equation}
p(X,\phi) = BX^n - V(\phi) \,,
\end{equation}
where $n$ is a positive integer (equal to 1 in the canonical case).
\item {Tachyon model}: The Tachyon  model was introduced in Refs.~\cite{Sen:2002nu,Sen:2002in}, and later studied in Refs.~\cite{Gibbons:2002md,Fairbairn:2002yp,Kofman:2002rh,Piao:2002vf}. Its Lagrangian is
\begin{equation}
\label{m:tachyon}
p(X,\phi) = -V(\phi) \sqrt{1-2 \lambda_{\rm s} X} \,,
\end{equation}
where the warp factor $\lambda_{\rm s}$ is a constant.
\item {DBI inflation model}: Its Lagrangian is given by
\begin{equation}
\label{m:DBI}
p(X,\phi) = - \frac{1}{f(\phi)} \sqrt{1-2 f(\phi) X} - V(\phi) \,,
\end{equation}
where we follow Ref.~\cite{Alishahiha:2004eh} and take the warp factor $f(\phi) \simeq \lambda_{\rm s}/\phi^4$ with $\lambda_{\rm s}$ constant. Some authors, e.g.\ Ref.~\cite{Silverstein:2003hf}, include an additional term $1/f(\phi)$ to cancel the leading-order term from expanding the square root. We have absorbed such a term into $V(\phi)$. Investigations of this model include Refs.~\cite{Peiris:2007gz,Lorenz:2008et,Powell:2008bi}; in particular Ref.~\cite{Bean:2007eh} contains a detailed study of one particular regime of this model.
\end{enumerate}
The sound speed is defined as
\begin{equation}
\css = \frac{\delta p}{\delta \rho} = \frac{\partial{p}/\partial X}{\partial\rho/\partial X}\,,
\end{equation} 
and one can show that for both non-canonical inflation models it is $\cs = \sqrt{1 - 2f(\phi)X}$, where the warp factor takes the unified form as $f=f(\phi)$. This factor is constant in Tachyon models, and a function of the scalar field in the DBI inflation models. 

We investigate the observables of interest, being the power spectrum $\Ps$, its spectral index $n_{{\rm s}}$, and the tensor--scalar ratio $r$, within the slow-roll approximation. Following Refs.~\cite{ArmendarizPicon:1999rj,Garriga:1999vw} (see also Refs.~\cite{Lorenz:2008et} and \cite{MRV} for more precise computations of the spectra), these are given by
\ba
\Ps &=& \frac{1}{8\pi^2} \frac{H^2}{\cs\epsilon} \label{mod-ps}\,,\\
\ns &=& \frac{d\ln\Ps}{d\ln k} = - (2\epsilon + \eta + s)  \label{mod-ns}\,, \\
r & = & 16 \cs \epsilon \label{mod-r} \,,
\ea
where the various small parameters are defined by
\be
\epsilon = -\frac{d\ln H}{Hdt} \spc \eta = \frac{d\ln\epsilon}{Hdt} \spc s = \frac{d\ln\cs}{Hdt} \label{mod-pms} \,.
\ee
In this paper, for each case we will focus on power-law potentials, either with general exponent $m$ or the simplest cases $V= \frac{1}{2} M^2 \phi^2$ and $V = \frac{1}{4}\lambda \phi^4$. Also we take the reduced Planck mass $\Mpl=1$ for convenience, for instance in the expression for $\Ps$.

We do not make an extensive discussion of non-gaussianity in this article, as the theory of non-gaussianity in these models is already well developed, and for instance has been applied to DBI models in the recent {\it Planck} analysis of Ref.~\cite{Ade:2013ng}. It is well known that non-canonical models are a way of generating detectable non-gaussianity, of equilateral shape. But this by no means implies that typical non-canonical models do; it takes quite some effort to obtain a non-gaussianity parameter above unity. By contrast, if one analyses simple potentials and kinetic terms, as we do in this article, typically the non-gaussianity remains small due to slow-roll suppression. This is true of all but one of our models, the exception being the DBI in the relativistic case where the constraint has already been provided in Ref.~\cite{Ade:2013ng}. Hence accurate calculations of the scalar and tensor power spectra, as provided here, are the only way to constrain these models at present.

\sect{The General Systematic Method}\label{sect-methods}

Here we propose a systematic method in the slow-roll regime to obtain and express the final solutions for observables, such as the power spectrum $\Ps$ and spectral index $n_{\rm s}$, as a function of e-folds $N$. 

\subs{General formula for the spectral index $n_{{\rm s}}$}\label{sect3}

We continue with the general Lagrangian $\L=p(\phi,X)$, noting for later use that any field redefinition to a new field $\varphi$, that is a function of $\phi$ and whose kinetic energy density is $\tX = \frac{1}{2} \partial_\mu \varphi \partial^\mu \varphi$, still leaves the Lagrangian as a general function $\tp(\varphi,\tX)$ and hence results in the same field equations in the new variables.

We write down the energy density of the universe, $\rho$, as usual \cite{ArmendarizPicon:1999rj,Garriga:1999vw}
\be\label{dens-eq}
\rho = 2 X p_{,X} - p \,.
\ee
The Friedmann equations are 
\be\label{frw-eq}
H^{2} = \frac{\rho}{3}  \spc \dot{H} = -X p_{,X} \,,
\ee
where we continue to use the convention $\Mpl = 1$. 

We now define a variable $u$ by
\be\label{def-u}
u \defeq \frac{1}{\epsilon}= \frac{H^{2}}{X p_{,X}} \,.
\ee
We will use Eqs.~(\ref{dens-eq}), (\ref{frw-eq}), (\ref{def-u}) and their derivatives to obtain the observables for the considered inflationary scenarios in the slow-roll regime.

Eq.~(\ref{frw-eq}) gives the continuity equation
\be\label{conti}
\rho' = 3(\rho + p)  = 6Xp_{,X} \,,
\ee
where $^{\prime}$ indicates derivative w.r.t.\ the e-folds $N$, which as usual are counted backwards from the end of inflation. According to Eqs.~(\ref{def-u}) and (\ref{conti}) we can write
\be\label{u-drr}
\frac{2}{u} = \frac{\rho'}{\rho} \,.
\ee
This compact form suggests that typically $\epsilon = 1/u \propto 1/2N$, in view of dimensional analysis. 

We wish to write the density of the Universe as
\be\label{rho-eps}
\rho =\rho(V(\phi),u) \,,
\ee
so that $X$ and $\dot{X}$ will play the role of simplification in recursion relations for $u$ and $\up$, etc. This form is general but useful, since it indicates that the considered quantities, such as the power spectrum, are determined by the potential and a small parameter $1/u$. The final result can be obtained by the perturbation method in terms of the small parameters coming from $u$ and its derivatives (here $\epsilon$, $\eta$, $\cs$, $s$). Once we find the solution for the quantity $u$ we can then obtain the potential via Eq.~(\ref{u-drr}), where it is a first derivative w.r.t.\ the scalar field $\phi$. 

To find an expression for $u$ without assuming a particular form of the Lagrangian, we differentiate Eq.~(\ref{def-u}) w.r.t.\ $N$ to obtain
\ba\label{duu-def}
\frac{\up}{u} &=& \frac{2}{u} - \frac{X^{\prime}}{X}\frac{1+\css}{2\css} - \frac{p_{,X\phi}}{p_{,X}}\phi^{\prime}  \,.
\ea
In deriving this equation, we have applied the relation $1/\css = 1 + 2 Xp_{,XX}/p_{,X}$. 

Differentiating Eq.~(\ref{u-drr}) w.r.t.\ $N$,
\be\label{duu-drr}
-\frac{\up}{u} = \frac{\rho''}{\rho'} - \frac{\rho'}{\rho} = \frac{2}{u} {\lrb \frac{\rho''\rho}{{\rho'}^{2}} - 1 \rrb}  \,,
\ee
and the relation $X=\frac{1}{2}H^{2}\phi^{\prime 2}$ gives
\be\label{dxx}
\frac{X'}{X} = 2 {\lrb \frac{\phi^{\prime\prime}}{\phi^{\prime}} + \frac{1}{u} \rrb} \,.
\ee
Equation~(\ref{dxx}) can be used to eliminate the field-dependent terms, such as $X'/X$ and $\phi^{\prime}/\phi$, which are present in Eqs.~(\ref{duu-def}) and (\ref{duu-drr}).

The quantity $\up$ has a clear meaning, namely
\be\label{eta-eps-sr}
\up = \frac{\eta}{\epsilon} \,. 
\ee
Once $u$ is obtained, we will have an explicit relation between $\eta$ and $\epsilon$. 

\ssubs{Exact formula for $u(c_{\rm s})$}

We now use Eq.~(\ref{conti}) to reformulate Eq.~(\ref{duu-def}) as,
\be\label{du-cs-v}
\up = 2 \left(1 - \frac{u}{2}\frac{p_{,X\phi} \phi^{\prime}}{p_{,X}}\right) - 3(1+\css)\left(1 - \frac{u}{2} \frac{\rho_{\phi}\phi^{\prime}}{\rho}\right)u  \,.
\ee
Equation (\ref{du-cs-v}) is derived from the field equation without any assumptions or model specification. It is general and has a quite symmetric form, from which we can get some descriptive results by inserting or approximating Eq.~(\ref{u-drr}). For example, qualitatively, if one approximates $\rho \sim V$, then for the following two models we will have
\bi
\item Canonical inflation\\ 
This type of inflation model has Lagrangian $\L = X - V(\phi)$. Then according to the equation above, we will have $\epsilon = 1/2N$, and $\eta = 2 \epsilon = 1/N$, due to
\be\label{two-cases-i}
p_{,X\phi} \equiv 0 \spc \frac{2}{u}  \simeq \frac{\rho_{\phi}\phi^{\prime}}{\rho} = \frac{V_{,\phi}\phi^{\prime}}{V}\,.
\ee
\item Tachyon models\\
The Lagrangian for this model is $\L = - V(\phi) \sqrt{1 - 2 \lambda_{\rm s} X}$, where $\lambda_{\rm s}$ is a constant. As in the analysis above, we find $\epsilon = {\rm const.}$ due to, 
\be\label{two-cases-ii}
\frac{p_{,X\phi}\phi^{\prime}}{p_{,X}} \equiv \frac{V_{,\phi}\phi^{\prime}}{V} \equiv \frac{\rho_{\phi}\phi^{\prime}}{\rho}  \simeq \frac{2}{u} \,.
\ee
\ei
The outcomes are  concise, but the approximate results of Eqs.~(\ref{two-cases-i}) and (\ref{two-cases-ii}) are only for qualitative understanding. We cannot make this approximation to eliminate any term in Eq.~(\ref{du-cs-v}). The evolution equation for $u$ is derived from the exact equation of motion, and if one wants to make an assumption on any term in Eq.~(\ref{du-cs-v}), one must also rederive the evolution equation for $u$. In the next subsection we address this issue. Therefore, although we have obtained this equation, it does not yet lead to a clear understanding for the observables for given model.

\ssubs{Predictions within the slow-roll approximation}

We need to derive a suitable equation for $u$.  In the following, we will apply a well-defined approximation scheme to derive results based on Eqs.~(\ref{def-u}), (\ref{conti}), (\ref{duu-def}), and (\ref{duu-drr}). In obtaining the observables, we use the variable $u$ and its  derivatives. 

In the slow-roll regime, the following equation is obtained after expanding Eq.~(\ref{conti})
\be\label{dxp-sr}
\rho_{,\phi} \phi^{\prime} \simeq 6Xp_{,X} \,.
\ee
Then Eq.~(\ref{u-drr}) can be written as
\be\label{rho-eps-loa}
\frac{2}{u}  = \frac{\rho_{,\phi} \phi^{\prime}}{\rho} \,,
\ee
Then the full Eq.~(\ref{du-cs-v}) will not applicable because it is derived from the full equation of motion for scalar field. We need to find the derivative of $X'$ from Eq.~(\ref{dxp-sr}) in order to get a similar equation for $\epsilon$ and $\cs$, by eliminating $X'/X$ in Eq.~(\ref{duu-def}). In view of this, we obtain the following equation,
\be
\frac{2}{u} \frac{\rho_{,\phi\phi} \rho}{\rho_{,\phi}^{2}} + \frac{\phi^{ \prime\prime}}{ \phi^{\prime}} =  \frac{X'}{X} \frac{1 + \css}{2\css} + \frac{p_{,X\phi}\phi^{\prime}}{p_{,X}} \label{dpp-sr} \,.
\ee
We also define a variable $\delta$,
\be
\delta \equiv \left( \frac{\rho_{,\phi\phi} \rho}{\rho_{,\phi}^{2}} - \frac{1}{2} \right)  \label{drr-sr} \,,
\ee
for later convenience. Assembling Eqs.~(\ref{duu-def}), (\ref{dxx}), (\ref{dxp-sr}), (\ref{dpp-sr}), and (\ref{drr-sr}) we eliminate $Xp_{,X}$, $\phi^{ \prime\prime}/ \phi^{\prime}$ and $X'/X$, but we keep the sound speed $\cs$ as it can be related to the variable $u$. After some effort, we can obtain the final result for Eq.~(\ref{duu-def}),
\be\label{ode-u-sr}
\up = 2 \lsb 1 - \delta(1+\css) \rsb    + \frac{p_{,X\phi}\phi^{\prime}}{p_{,X}} u \css  \,.
\ee
As we have noted in Eqs.~(\ref{two-cases-i}) and (\ref{two-cases-ii}), the last term in the above equation can now be approximated and then the whole equation can be simplified and solved. 

\subs{Predictions for two classes of models}

Using Eq.~(\ref{ode-u-sr}) we consider some classes of inflation models. Unless explicitly stated, the following discussion will consider a monomial potential of the form $V \propto \phi^{m}$. This will provide a constant $\delta =1/2 - 1/m$, independent of the field value itself.

\ssubs{Variable separable class}

This class of inflation models includes two subclasses: Sum-Separable Models ({\bf SSMs}) and Product-Separable Models ({\bf PSMs}).
\bi
\item[$\star$] \ul{{\bf SSMs}:}  Here we have separate terms for $\phi$ and $X$ which are added together, for example $\mathcal{L} = BX^{n} - A\phi^{m}$.\footnote{Actually when the kinetic term consists of a single monomial term, any constant prefactor can readily be set to unity without loss of generality by a field rescaling.} In this class, similarly to Eq.~(\ref{two-cases-i}), we have the following useful relations
\be
\rho(V(\phi),u) = \frac{V(\varphi)}{1-1/3u} \spc p_{,X\phi} \equiv 0 \spc {\css} \equiv \frac{1}{2n-1}  \,,
\ee
which can be substituted into Eq.~(\ref{ode-u-sr}). Then we can obtain a compact form, recovering the results we found for this model in Ref.~\cite{Li:2012vt}: 
\be\label{can-du-N-sr}
\up = 2 \frac{\beta}{m}   \spc \beta = \frac{(n-1)m + 2n}{2n-1}  \,.
\ee
{\it Solution for $u$:} According to the equation above, the solution is explicitly obtained as
\be\label{can-eps-N-sr}
\epsilon = \frac{m}{\beta} \frac{1}{2N}  \,.
\ee 
We have a model independent $\eta$ according to Eqs.~(\ref{eta-eps-sr}), (\ref{can-du-N-sr}) and (\ref{can-eps-N-sr}),
\be\label{eta-N-sr}
\eta = 2 \frac{\beta}{m} \epsilon \equiv \frac{1}{N}  \,,
\ee
under the linear assumption by considering only the leading-order small parameter, such as $\epsilon$. We see that the parameter $\epsilon$ depends on the exponents $m$ and $n$ of the given model, while the parameter $\eta$ does not.

{\it Power spectrum $\Ps$, spectral index $n_{{\rm s}}$, and tensor-to-scalar ratio $r$:} Now that we have obtained the solution for $u$ expressed by Eq.~(\ref{can-eps-N-sr}), we can use Eqs.~(\ref{dxp-sr}), (\ref{rho-eps-loa}), and (\ref{can-du-N-sr}) to find the scalar potential $V=A\phi^m$ in terms of e-folds $N$. We recall that Eq.~(\ref{dxp-sr}) gives the slow-roll prediction
\be
\Vp = 6^{1-n}n B \frac{V^{n-1}}{(\beta N)^{2n}} {\lrb \frac{V}{A} \rrb}^{2n/m} \,,
\ee
and Eq.~(\ref{rho-eps-loa}) gives
\be
\frac{\r^{\prime}}{\r} \simeq \frac{\Vp}{V} = \frac{2}{u} \,.
\ee
Finally we obtain the potential $V$ as
\be\label{can-V-N-sr}
V = {\lrb \frac{m6^{n-1}}{n} \, \frac{A^{\frac{2n}{m}}} {B} \rrb}^{\frac{m}{\beta}\frac{1}{2n-1}} {(\beta N)}^{m/\beta} \,.
\ee
Therefore we can now write down the power spectrum from Eq.~(\ref{mod-ps}) and its spectral index from the Eq.~(\ref{mod-ns}),
\ba
\Ps &=& \frac{\sqrt{2n-1}}{12\pi^{2}} \frac{\beta}{m} \times {\lrb \frac{m\beta^{2n-1}6^{n-1}}{n} \, \frac{A^{\frac{2n}{m}}} {B}  \rrb}^{\frac{m}{\beta}\frac{1}{2n-1}} \times N^{\frac{m}{\beta}+1}  \label{ps-ssm} \,,\\
\ns &=& -2 \left(1+\frac{m}{\beta} \right) \times \frac{1}{2N} \label{ns-ssm}\,,\\
r &=& 16\frac{m}{\beta} \frac{1}{2N} \label{r-ssm}\,.
\ea

\item[$\star$] \ul{{\bf PSMs}:} These take the form $\mathcal{L} = -K(X)V(\phi)$ where both $K(X), V(\phi)>0$, the Tachyon being an example. We will still have the following relations, analogous  to the case in Eq.~(\ref{two-cases-ii}),
\be
\rho_{,\phi} \propto V_{,\phi}  \spc \frac{p_{,X\phi}\phi^{\prime}}{p_{,X}} \equiv \frac{V_{,\phi}\phi^{\prime}}{V} = \frac{2}{u} \label{tachyon-u-phi}  \,,
\ee
which leads Eq.~(\ref{ode-u-sr}) to be,
\be\label{tachyon-du-N-css-sr}
\up = 2(1 - \delta) (1+\css) \,.
\ee
Here we have an implicit function, the sound speed $\css = \css(u)$. To get an explicit result for model observables, such as $\Ps$ and $\ns$ via this differential equation (\ref{tachyon-du-N-css-sr}), we first need to specify the particular form of $\css$ in terms of $u$. In general it is not straightforward to solve this equation due to the undetermined sound speed, but this is possible for the Tachyon Models as presented in Section~\ref{app-to-sect-tachyon}.
\ei

\ssubs{A more general ansatz}

Not all Lagrangians are sum or product separable, of course; the Tachyon is product-separable but the DBI case is neither. To set up formalism to deal with the latter, we consider the more general Lagrangian 
\be
\label{e:330}
\L = - \left( W(\phi) K(X) + U(\phi) \right) \,.
\ee
This contains the SSM and PSM as special cases, respectively $W$ constant and $U \propto W$. In its conventional form, Eq.~(\ref{m:DBI}), the DBI model does not take this form, but we will show below that it can be written in this form via a field redefinition. Although we will not undertake a general study for the above ansatz, due to the complexity of the analysis, we will use the DBI case to illustrate our procedure on Lagrangians of this class.

\sect{Application of the Systematic Method to Two Models}\label{sect-twomods}
\subs{Tachyon models}\label{app-to-sect-tachyon}
We now consider the Tachyon model given by Eq.~(\ref{m:tachyon}) and apply  Eq.~(\ref{tachyon-du-N-css-sr}). At this point we don't have to impose any simplification since the Lagrangian already takes the product-separable form. However, we will implement a field-redefinition approach to rederive these results in a different way in Section~\ref{canon-tach}.

For this type of model, the relation between $u$ and the sound speed $\cs$ is
\be\label{tachyon-css-u}
\epsilon = \frac{1}{u} = \frac{3}{2}(1-\css) \,,
\ee
while the energy density $\rho$ is
\be
\rho(V(\phi),u) = \frac{V(\phi)}{\sqrt{1-2/3u}} \simeq V(\phi) \,.
\ee
Then we can solve for $u$ according to  Eq.~(\ref{tachyon-du-N-css-sr}), which can be written as
\be
\up = 4\mu \left(1 - \frac{1}{3u}\right) \spc \mu  = 1 - \delta = \frac{1}{2} + \frac{1}{m} \,.
\ee
The solution for $\epsilon = 1/u$ is therefore
\be\label{tachyon-u-N-sr}
N = \frac{1}{4\mu} {\left[ \frac{1}{\epsilon} + \frac{1}{3}\ln{\left(\frac{3}{\epsilon} - 1\right)} \right]}  \,.
\ee
The solution can be inverted using the Lambert $\mcl W$ function\footnote{Two useful properties of the Lambert $\mcl W$ function \cite{lambert-wiki} are $\mcl{W}(e) = 1 \label{W-e-1}$
and its asymptotic behaviour for any real $x \geq e$ \cite{lambertw-HH},
$$
L_1 - L_2 + \frac{1}{2}L_3\leq \mcl W(x) \leq L_1 - L_2 + \frac{e}{e-1}L_3 
$$
where $L_1 = \ln{x} ,\; L_2 = \ln\ln{x} ,\; L_3 = L_2/L_1$.}
so we will have,
\be
u = \frac{1}{\epsilon} = 1 + \mcl W(e^{x-1}) \spc x = 12\mu N \label{tachyon-u-eps-W} \,.
\ee
[The Lambert function is not really necessary here, but we use this method as it is required in later parts of this article.]
As $\mu$ is of order one, $x \gg 1$ for any $N$ of interest, which means that $\epsilon \ll 1$ will always hold. Then we can obtain the following relations
\ba
\epsilon &=& \frac{1}{2\mu}\frac{1}{2N} \,,  \label{tachyon-u-N-sr-close}\\
\eta = 4\mu \epsilon \left(1 - \frac{\epsilon}{3}\right) &=&  \frac{1}{N} {\lrb 1 - \frac{1}{6\mu} \frac{1}{2N} \rrb} \simeq \frac{1}{N}\,.
\ea
To find the power spectrum, we need to find the relation of $\phi$ to $N$. To achieve this, we combine Eq.~(\ref{dxp-sr}) with Eqs.~(\ref{tachyon-u-phi}), (\ref{tachyon-css-u}), and (\ref{tachyon-u-N-sr-close}) and find the potential $V$ in terms of $N$ and the model parameters $A,\, \ls$,
\be
V = \kappa N^{1/2\mu} \spc  \kappa = {\lrb 2m^{2}\mu \frac{\cs A^{2\mu-1}}{\ls} \rrb}^{1/2\mu} \,.
\ee
According to Eq.~(\ref{mod-ps}) the scalar power spectrum is
\be\label{tachyon-ps-N}
\Ps = \frac{1}{12\pi^{2}}\cs^{1/2\mu -1} \times{\lrb m^2 (2\mu)^{1+2\mu} \frac{A^{2/m}}{\ls} \rrb}^{1/2\mu} \times N^{1+1/2\mu}\,.
\ee

The spectral index receives contributions from the last term and also from the time variation of the sound speed $\cs$, but we will now see that the latter term does not contribute at lowest order in slow-roll. To check this, we need $s$ which is
\be
s = -\frac{\cs^{\prime}}{\cs} = \frac{1-\css}{2\css} \frac{X^\prime}{X} \,.
\ee
Combining with $\up$ from Eq.~(\ref{ode-u-sr}), we have
\be
\up = -\frac{1+\css}{2\css} \frac{X^\prime}{X} u \,,
\ee
and therefore
\be\label{tachyon-s-N-sr}
s = \eta \frac{- \epsilon/3}{1 - \epsilon/3} = -\frac{4}{3} \mu \epsilon^{2} \spc  \frac{X^\prime}{X} = 4(\delta-1)\epsilon\css \,.
\ee
The spectral index is then
\be\label{tachyon-ns-sr}
\ns = -(2\epsilon + \eta + s)  = - (2+4\mu)\epsilon + \frac{8}{3}\mu\epsilon^{2} \,.
\ee
Retaining only the lowest-order terms in slow-roll, as required for consistency as only those have been included throughout, we have the spectral index $n_s$ and $r$ as
\ba
\ns &=& -\frac{2m+2}{m+2}\frac{1}{N} \label{tachyon-ns-N-est}\,,\\
r &=& \frac{8\cs}{\mu}\frac{1}{2N} \simeq \frac{8m}{m+2}\frac{1}{N} \label{tachyon-r-N-est}\,.
\ea
This indicates that the spectral index is always red tilted, $n_{{\rm s}} < 1$, in Tachyon models. The number of e-folds between observable perturbation generation and the end of inflation, usually taken to be $N \simeq 50$ \cite{Liddle:2003as}. So take an example for quadratic potential where $m=2$, we will have $n_s=0.97$ and $r=0.08$ at pivot $N_*=50$.

\subs{DBI inflation models}\label{app-to-sect-dbi}

The DBI action, as already given in Eq.~(\ref{m:DBI}), is
\be\label{m:DBI-2}
p(\phi,X) = -\frac{1}{f(\phi)} \sqrt{1 - 2f(\phi) X} - V(\phi) \,,
\ee
where the sound speed $\cs = \sqrt{1 - 2f(\phi) X}$ and the warp factor is $f(\phi) = \ls/\phi^4$. If the last term $V$ is zero the model reduces to the Tachyon model with a constant potential, but we are not interested in this case here; instead we are going to discuss a more general case in the following subsections.

\ssubs{Field redefinition}

To proceed with our investigation using the method of the previous sections, and in view of  simplifying the later calculations, we apply a field redefinition $\varphi = 1/\phi$ to the DBI action~(\ref{m:DBI}). A variant of this technique will also be used in studying the Tachyon model in Section~\ref{sect-implement}. However, we will now focus on the current case, where the Lagrangian for DBI inflation with a potential $V = A\phi^m$ becomes
\ba
\varphi &=& \frac{1}{\phi} ,\;W(\varphi) = \frac{1}{\varphi^{4}} ,\; \tV(\varphi) = A\varphi^{4-m}\,, \label{dbi-phi-vpahi}\\
\L &=& - W(\varphi) {\lrb \frac{1}{\ls}\tilde{\cs} + \tV(\varphi) \rrb} \label{dbi-vphi-L} ,\, \tilde{\cs} = \sqrt{1 - 2\ls \tX} \label{dbi-cs-red-tX} \,.
\ea
The notation $\tX$ stands for the kinetic term after applying the field redefinition. With this new field definition, the DBI action falls in the class defined by Eq.~(\ref{e:330}).

The energy density $\r$ is
\be
\r = W(\varphi) {\lrb \frac{1}{\ls \cs} + \tilde{V}(\varphi) \rrb} = \frac{3}{2}\frac{W}{\ls} \frac{1-\css}{\cs} u \,.
\ee
We have a sound speed (for later convenience, we use the same notation $\cs$ for sound speed instead of $\td\cs$) which is of the same form as for the Tachyon. The above action in Eq.~(\ref{dbi-cs-red-tX}) will induce an equation of motion,
\be\label{dbi-vphi}
\frac{\ddot\varphi}{\cs^{2}}  + 3H \dot\varphi + \left[\frac{W^{\prime}}{W} \frac{1}{\ls} + \cs\tilde{V}\left(\frac{\Wp}{W} + \frac{\tilde\Vp}{\tilde{V}} \right) \right] = 0 \,.
\ee
Rearranging Eq.~(\ref{dbi-vphi}) it can be reformulated as
\be\label{dbi-vphi2}
\frac{\ddot\varphi}{\cs^{2}}  + 3H \dot\varphi + \frac{W^{\prime}}{\ls W} \left[ 1 + {\ls\cs}\tilde{V}\left(1+ \frac{\tilde\Vp}{\tilde{V}}\frac{W}{\Wp} \right) \right] = 0\,,
\ee
where the Hubble rate is
\be\label{dbi-vphi-H}
H^{2} = \frac{u}{2}\frac{W}{\ls}\frac{1-\css}{\cs} \,.
\ee
Also Eq.~(\ref{dbi-vphi2}) suggests the same form as the Tachyon, if the potential $V$ is constant.

\ssubs{Predictions for the quartic potential}\label{sect-eval-dbi-lp4}

In this section we will consider the quartic potential $V = \frac{\lambda}{4}\phi^4$, which will provide a simple form of differential relation in Eq.~(\ref{tachyon-du-N-css-sr}) for $u = 1/\epsilon$ and its first  derivative. We can write
\ba
\r = \frac{W(\varphi)}{\ls\cs} ( 1 + \alpha \cs ) &\spc& \alpha = A\ls = \frac{\ls\lambda}{4} = {\rm const} \,, \\
\delta = \frac{\Wff W}{\Wf^{2}} - \frac{1}{2} \equiv \frac{3}{4} &\spc& \epsilon = \frac{3}{2} \frac{1 - \css}{1 + \alpha \cs} \label{dbi-eps-cs} \,.
\ea
Unlike the case of Tachyon models, for the quartic potential model in DBI inflation we have a constant $\delta$. We now have the approximate equation for relation (\ref{dxp-sr}),
\be\label{dbi-sr-lp4}
\frac{2}{u} = \frac{\r_{,\varphi}\varphi^{\prime}}{\r} \equiv \frac{W_{,\varphi}\varphi^{\prime}}{W} = \frac{p_{,\tX\varphi} \varphi^{\prime}}{p_{,\tX}}\,.
\ee
Also, we can readily solve Eq.~(\ref{dbi-sr-lp4}) so that we can write down the power spectrum $\Ps$ and its spectral index $n_{{\rm s}}$. This is simplified because of the field redefinition, in contrast to the conventional treatment where one could not get a slow-roll solution for $\Ps$ etc. According to Eq.~(\ref{dxp-sr}) and Eqs.~(\ref{dbi-vphi2}) and (\ref{dbi-sr-lp4}) we have\footnote{One can just expand Eq.~(\ref{dbi-eps-cs}) to obtain this relation, but for consistency of our treatment we present the same procedure as followed in the previous sections.}
\be
3X = \epsilon \frac{1+\alpha\cs}{\ls}\,,
\ee
which then solves $\varphi$ and the redefined warp factor $W(\varphi)$,
\be
W = 64\css \frac{1}{\epsilon^2} \label{dbi-W-cs-eps-sr} \,.
\ee
Then we have $\Ps$ as
\be
\Ps = \frac{8}{3\pi^2} \frac{1+\alpha\cs}{\ls} \frac{1}{\epsilon^3} \label{dbi-Ps-cs-eps-ls-1-sr}\,.
\ee
Now we notice that there is another relation for $\epsilon$ in Eq.~(\ref{dbi-eps-cs}), so the above expression can be written in another equivalent form,
\be
\Ps = \frac{4}{\pi^2} \frac{1-\css}{\ls} \frac{1}{\epsilon^4} \label{dbi-Ps-cs-eps-ls-2-sr} \,.
\ee
These formulae may be used interchangeably. Considering the two asymptotic limits for the sound speed $\cs=0$ or $1$, either of the above leads to the prediction for the power spectrum. In the limit $\cs\sim1$, Eq.~(\ref{dbi-Ps-cs-eps-ls-1-sr}) straightforwardly indicates the leading dependence for $\Ps$ of order $1/\ls\epsilon^3$, while in the case  $\cs \raw 0$ we may use the second expression for the power spectrum $\Ps$ by Eq.~(\ref{dbi-Ps-cs-eps-ls-2-sr}). This alternative equation is more useful in some cases. For example, providing no knowledge about the model parameter degeneracy which is encoded in $\alpha$, we cannot obtain good insight from Eq.~(\ref{dbi-Ps-cs-eps-ls-1-sr}) in the cases where the limits $\cs \raw 0$ or $\alpha \raw 0$ are approached. On the contrary, Eq.~(\ref{dbi-Ps-cs-eps-ls-2-sr}) will tell us the prediction in spite of $\cs$ or the degenerate combination $\alpha$. And in the limit of $\cs \sim 0$, we will obtain the usual prediction for the power spectrum $\Ps = 4/\pi^2 \ls \epsilon^4$ when the inflaton field moves relativistically in DBI inflation.

We would however like to go further in obtaining the expression for $\Ps$ for the reason that in either of the expressions above, which have two or three small parameters, it remains unclear where the model degeneracies lie. Therefore, as in previous sections, we are going to find an expression for $u$ in terms of e-folds $N$. To do this, following Eq.~(\ref{tachyon-du-N-css-sr}), we obtain a similar equation for $u$ 
\be\label{dbi-du-cs-sr}
\up = \frac{1}{2} (1 + \css) \,.
\ee
However, as $u$ is a rational function of $\cs$ (see Eq.~(\ref{dbi-eps-cs})), we will use another approach to obtain the solution for $u$.

Assembling all of Eqs.~(\ref{duu-def}), (\ref{dbi-cs-red-tX}), and (\ref{dbi-sr-lp4}), we will have the ODEs below 
\ba
\frac{\up}{u} &=& - \frac{{\tX}^{\prime}}{2\tX} \frac{1+\css}{\css}  - {\lrb \frac{2}{u} - \frac{p_{,\tX\varphi}\varphi^{\prime}}{p_{,\tX}} \rrb} \,,\label{d1} \\
\frac{\cs^{\prime}}{\cs} &=& -\frac{1-\css}{\css}\frac{\td{X}^{\prime}}{2\td X} = -s  \label{dbi-sr-dcs}\,.
\ea
We turn to investigate the evolution equation (\ref{dbi-sr-dcs}) for the sound speed, whose evolution is very simple due to the field redefinition to $\varphi$-field space. 

We can solve the relation for $\td{X}^{\prime}/\td X$ by using Eqs.~(\ref{dbi-sr-lp4}), (\ref{dbi-du-cs-sr}), and (\ref{d1})
\be
\frac{\td{X}^{\prime}}{\td X} = -\frac{\css}{u} \,,
\ee
then we get the evolution equation for sound speed $\cs$,
\be\label{dbi-cs-N-ode}
\cs^{\prime} = \frac{3}{4} \frac{(1-\css)^{2}}{1 + \alpha \cs} \cs \,.
\ee
The solution is obtained as follows
\be\label{dbi-sr-N-eps-cs}
N = \frac{1}{\epsilon} + \frac{2}{3} {\lrb  \ln{\frac{\css}{1 - \css}} + \frac{\alpha}{2} \ln{\frac{1 + \cs}{1 - \cs}} \rrb} \,.
\ee
We are interested in the predictions in the different limits of the sound speed $\css$, although we cannot explicitly invert the general results.
\bi
\item[$\star$] {\it Non-relativistic case}\\
It is not generally possible to invert the relation for $\epsilon$ and the e-folds $N$ since $\alpha$ is an important parameter. Then we will not have the full expression for the scalar power spectrum. However, in the non-relativistic case the condition $\css \sim 1$ can be applied such that the latter terms are negligible. So we will have the approximate solution for $\epsilon \simeq 1/N$. Then by this relation, we obtain the redefined scalar field $\varphi$,
\be\label{dbi-sr-W-N-cs}
W(\varphi) = \tau N^{2} \spc \tau = 64\css \,.
\ee
Finally we have the scalar power spectrum by Eq.~(\ref{mod-ps}) as\footnote{To determine $\tau$, we applied Eq.~(\ref{rho-eps-loa}). It is a bit tricky to determine the constant $\tau$ in this case. This relation is for $\css \raw 1$; one should replace $1-\css$ with (1 + $\alpha\cs$) via the definition of $\epsilon$. Then one finds this proportionality coefficient is exactly $64\css$.}
\be\label{dbi-sr-ps}
\Ps \simeq \frac{8}{3\pi^2} \frac{1+\alpha}{\ls} N^3 \propto {\lrb A + \frac{1}{\ls} \rrb} N^3 \,.
\ee 
Therefore we can obtain the spectral index
\be\label{dbi-sr-ns}
\ns \simeq -\frac{3}{N} \,.
\ee
This relation can also be derived from the differential system. We can evaluate the two small parameters according to Eqs.~(\ref{eta-eps-sr}), (\ref{dbi-du-cs-sr}), and (\ref{dbi-sr-dcs}),
\be
\eta = \frac{1+\css}{2}\epsilon \spc s = -\frac{1-\css}{2} \epsilon \,,
\ee
to obtain the spectral index which is then presented as,
\be\label{dbi-sr-ns-eps-cs}
\ns = - (2\epsilon + \eta + s) = - (2 + \css) \epsilon \,,
\ee
in terms of $\epsilon$ and $\cs$. From Eq.~(\ref{dbi-sr-ns-eps-cs}) we find that for a quartic potential the spectrum is always red tilted, since $2+\css>0$ is always satisfied regardless of the warp strength or the mass of the scalar field. Also if the DBI scalar field rolls asymptotically in the manner of a canonical field in which $\css \raw 1$, then we have $n_{s} \raw 1 - 3/N_{*} = 0.94$ at $N_{*}=50$. This result is obviously recovered due to the canonical-like inflation with a quartic potential.

One should note that the relation in Eq.~(\ref{dbi-sr-ns-eps-cs}) does {\em not} require any limit for $\cs$. It can also be applied when $\css \ll 1$, but if that is the case we should modify the value for $\epsilon \simeq 2/N$. We will see this below.

\item[$\star$] {\it Relativistic case}\\
The behaviour of DBI inflaton can be relativistic, which corresponds to the limit $\css \ll 1$. In this case, we can also use the method from this section. Unlike the case of $\css \raw 1$ we only need to approximate $\epsilon \sim 2/N$.\footnote{This relation can be obtained either from Eq.~(\ref{dbi-cs-N-ode}) by inserting $\css\raw0$, or using limiting properties for the third term in Eq.~(\ref{dbi-sr-N-eps-cs}) which is $\log{\frac{1+\cs}{1-\cs}} \raw 2\cs$. In this limit, we will infer another relation for $\alpha\cs\epsilon \simeq 3/2$.}

Therefore, we can write down the relation for the redefined scalar field $\varphi$, the scalar power spectrum and its spectral index as follows
\ba
W = \xi N^4 &\spc& \xi \simeq {\lrb \frac{3}{\alpha} \rrb}^2 \label{dbi-rel-w-N} \,, \\
\Ps &=& \frac{1}{4\pi^2} \frac{1}{\ls} N^4 \label{dbi-rel-Ps-N} \,, \\
\ns &=& -\frac{4}{N} \label{dbi-rel-ns-N}\,.
\ea
To derive the proportionality constant in Eq.~(\ref{dbi-rel-w-N}), we have used the speed limit relation $\ls{\dot\varphi}^2 \sim 1$. While deriving Eqs.~(\ref{dbi-rel-w-N}) and (\ref{dbi-rel-Ps-N}), we have also applied the relation $\alpha\cs \epsilon \simeq 3/2$ for this relativistic case. These relations can be obtained by inserting $\css \ll 1$ into Eq.~(\ref{dbi-eps-cs}). The sound speed in this scenario is constrained by non-gaussianity to be greater than about 0.07 \cite{Ade:2013ng}.
\ei
Since we can have slow-roll solutions for $\epsilon$ in each case, we can just substitute either solution to the power spectrum into Eq.~(\ref{dbi-Ps-cs-eps-ls-1-sr}) or Eq~(\ref{dbi-Ps-cs-eps-ls-2-sr}). Eventually we will have the predictions as above. We can see both predictions for the power spectrum for DBI inflation with quartic potential are valid. However, the first one (\ref{dbi-Ps-cs-eps-ls-1-sr}) will reduce to the conventional canonical inflation with quartic potential in the non-relativistic limit, while Eq.~(\ref{dbi-Ps-cs-eps-ls-2-sr}) will give the prediction for it in the relativistic limit.
%

\bfg[t]
\centering
\includegraphics[width=0.9\textwidth]{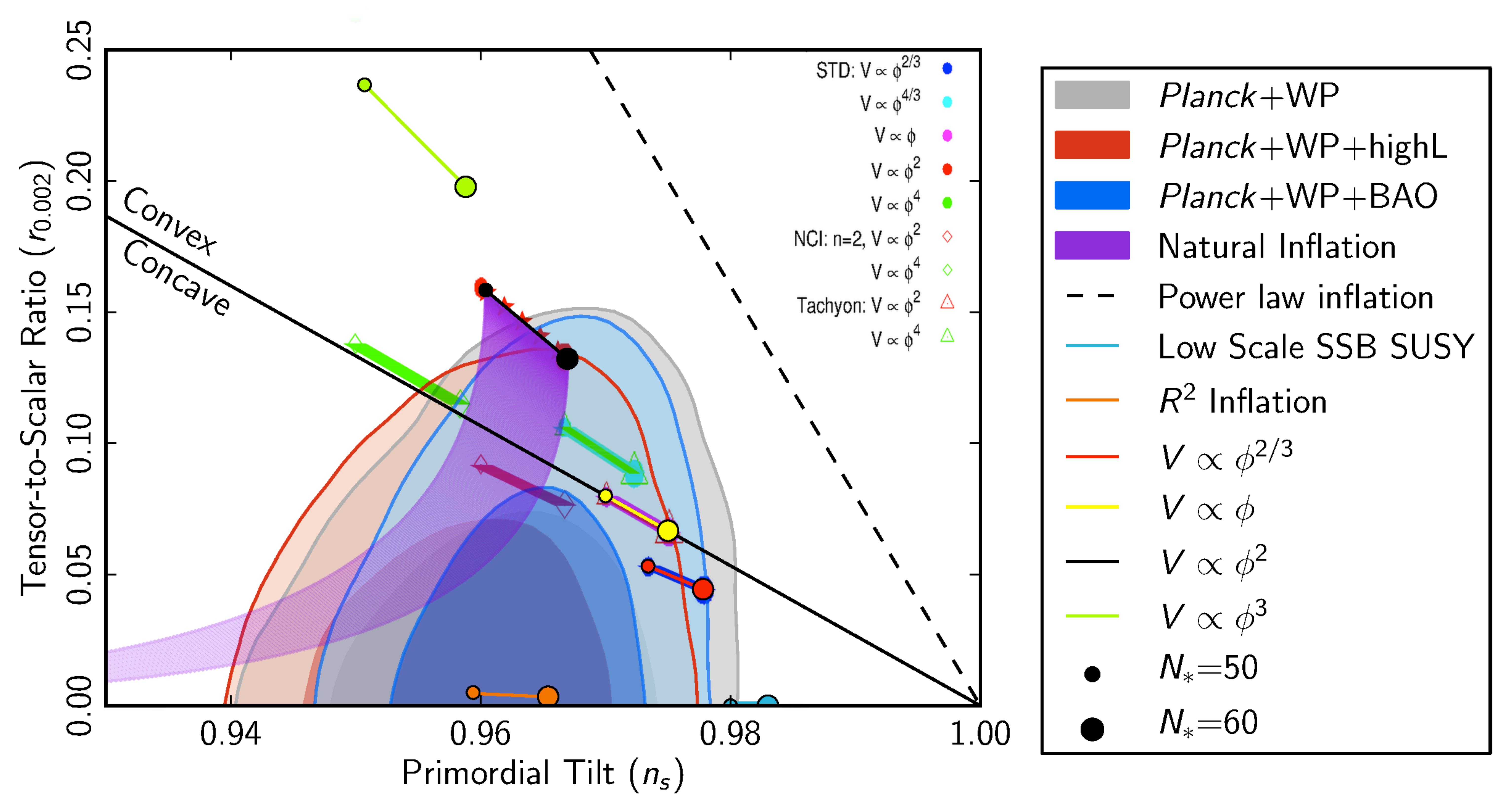}
\caption{Slow-roll predictions for several inflation models alongside constraints from {\it Planck} and other probes. [Based on an image from Ref.~\cite{Ade:2013uln}, original image credit ESA/{\it Planck} Collaboration.]}
\label{fig:all-3-nsr}
\efg

\subs{Comparison to {\it Planck} constraints}

We have studied several models and derived their predictions within the slow-roll approximation. Now we present these predictions by visualising them against the latest constraints on inflation models compiled by the {\it Planck} collaboration \cite{Ade:2013uln}.
The {\it Planck}+WP observational data favours a concave potential for viable canonical inflation models and limits the tensor-to-scalar ratio below $r=0.11$ at 95\% confidence. In Figure \ul{\ref{fig:all-3-nsr}} we can see that for canonical inflation models with the polynomial potential $V\propto \phi^m$, the $m=3,4$ cases are ruled out by the observational requirements, while potentials with $m=2/3$, $1$, $4/3$ and $2$ are within 95\% confidence, though at $N_*=50$ the quadratic potential ($m=2$) lies outside the 95\% confidence region \cite{Ade:2013uln}.

This plot shows two of our models. The first is the NCI model where the Lagrangian has the form $\L = BX^2 - V(\phi)$, where both quadratic potential (red diamond) and quartic potential (green diamond) are located within the observationally-permitted region. The second is Tachyon inflation; the figure shows that the quadratic potential (red triangle) is well within the permitted region, while the quartic potential is marginal with only the large $e$-foldings case $N_* = 60$ lying within the 95\% confidence region.

The polynomial potential in NCI-$X^2$ models and Tachyon inflation model can be considered as reshaped by the non-canonical term, and these reshaped potentials can correspond to standard inflation models. In $X^2$ models the quadratic potential is reshaped to $m\in(2/3,1)$ and the quartic potential to $m\in(1,4/3)$. In Tachyon models, the quadratic potential is reshaped to $m=1$ and the quartic potential to $m=4/3$. This reshaping in Tachyon models can be found in the following section.

One may infer from the figure that if the inflaton rolls in a polynomial potential $V\propto \phi^m/m$ with higher $m$, the higher-order kinetic term $X^n$ in NCI models may be supported by current observational datasets such as {\it Planck} and BAO. These models can potentially give significant non-gaussianity while satisfying power spectrum constraints, since large non-gaussianity occurs via a small sound speed $\cs$, which the NCI-$X^n$ models can provide with $\css = 1/(2n-1)$.

\sect{Implementation Methods for Particular Cases}\label{sect-implement}

For some particular models, there may be more direct treatments. For example for Tachyon models, we can take advantage of a field redefinition, along with the assumption $\epsilon \ll 1$ which in turn implies that $\cs \raw 1$. This method leads the same prediction as those obtained in Section \ref{sect-twomods}, as we show next. For DBI inflation with a quadratic potential, the methods in Section \ref{sect-twomods} require a lot of effort in order to obtain the observables, but initially applying a slow-roll assumption gives results for the model predictions. The following methods can be considered as an implementation of the systematic approach in Section \ref{sect-methods}. 

\subs{Field redefinition in Tachyon models}\label{canon-tach}

A more direct approach to the observables is possible for Tachyon models, as a combination of the slow-roll approximation and a field redefinition allows the theory to be approximated by a canonical Lagrangian.
Starting from
\be \label{tachyon}
P(X,\phi) = -V\sqrt{1-2\ls X},
\ee
and making the slow-roll assumption $\ls X \ll 1$, we can approximate it by a different non-canonical model with Lagrangian
\be \label{sim-tachyon}
P(X,\phi) \simeq V\ls X - V.
\ee
This can be transformed to a canonical action if we can find a new field $\varphi$, with corresponding kinetic energy $\tilde X = \frac{1}{2} \partial_\mu \varphi \partial^\mu \varphi$, such that $\tilde X= V\ls X$. This can be done in principle for any well-defined potential, though not always analytically. Again focussing on the monomial potential $V(\phi)=A\phi^m$, the required transformation is
\be
\varphi = \frac{2\sqrt{A\ls}}{m+2}\,\phi^{1+m/2} \,,
\ee
so that we have $V(\varphi)$ as,
\be \label{norm-non}
V(\varphi) = \varphi^{\frac{2m}{m+2}} \left(\frac{m+2}{2}A^{\frac{1}{m}} \ls^{-\frac{1}{2}} \right)^{\frac{2m}{m+2}}\,.
\ee
By completing the above transformation, then the original Lagrangian (\ref{tachyon}) can be rewritten within this approximation as
\be
P(X,\phi) \raw \tilde P(\tilde X, \varphi) = \tilde X - \tilde A  \varphi^{\frac{2m}{m+2}} \,,
\ee
where the normalisation coefficient became
\be
\tilde A = \left(\frac{m+2}{2} \right)^{\frac{2m}{m+2}} \left(A^{1/m} \ls^{-1/2} \right)^{\frac{2m}{m+2}} \,.
\ee
The usual results for canonical inflation with  $V(\varphi) \propto \varphi^\alpha$ then apply, giving a spectral index and tensor--scalar ratio of \cite{LL92}
\begin{eqnarray}
n_{\rm s} -1 & = & - \frac{2+\alpha}{2N} = - \frac{2m+2}{m+2} \frac{1}{N}  \,,\\ 
r & = & \frac{4\alpha}{N} = \frac{8m}{m+2} \frac{1}{N} \,.
\end{eqnarray}
These match the results we found in Section~\ref{app-to-sect-tachyon}.

We therefore conclude that, provided our slow-roll assumption holds, a power-law potential in a Tachyon  model behaves as if it were a canonical model but with a different power which is smaller (and indeed never bigger than 2). For example, the quadratic potential $m=2$ rescales to a linear potential $V(\varphi) = 2\sqrt{A/\ls}\,\varphi$, while the quartic potential $m=4$ rescales to
\be 
V(\varphi)=\left(\frac{3A^{1/4}}{\sqrt{\ls}} \right)^{4/3}\varphi^{4/3}\,.
\ee
The upper limit for the rescaled potential is the quadratic type $V(\varphi)= \hat A \varphi^{2}$, which in canonical inflation is well explored and permitted by present data including that from the {\it Planck} satellite \cite{Ade:2013uln}. Consequently, as long as the slow-roll assumption made at the start is valid, we expect this potential to match the observational data for any value of the power-law, unlike in the canonical case.

These results explain the degeneracy between the potential normalization and $\ls$ found numerically in Ref.~\cite{Devi:2011qm}, and also confirmed in some numerical analysis we have undertaken ourselves that we display later. As the spectral index and tensor--scalar ratio match observations, the only parameter tightly constrained by observations is the potential normalization in the $\varphi$ representation. Hence we predict a perfect degeneracy $A \propto \ls$ in the quadratic case and $A \propto \ls^2$ in the quartic case.

We can also now check the validity of the slow-roll assumption made to obtain these solutions. Within the canonical frame, it is well known that the slow-roll approximation $\tilde{X} \ll V(\varphi)$ is valid. Hence $V\ls X \ll V$, and hence $\ls X \ll 1$ as required for our original approximation in the Lagrangian. This shows the self-consistency of the slow-roll approximation we have deployed. Nevertheless, it remains possible that there are other observationally-valid solutions that do not obey the slow-roll condition; investigation of this would require a numerical analysis.

\subs{DBI inflation with a quadratic potential}\label{sect-dbi-m2p2-sra}

We now carry out a similar procedure for DBI inflation with a quadratic potential. According to Eq.~(\ref{dbi-vphi}), in which the first term is negligible compared to the second term (see Appendix~\ref{dbi-sra-validity}), we can obtain an approximate slow-roll equation,
\begin{equation}
3H \dot\varphi \simeq - \frac{W^{\prime}}{W} \frac{1}{\lambda_{\rm s}} \left( 1 + \frac{m}{4} \alpha c_s \varphi^{4-m} \right) \spc \alpha = A\ls \,. \label{sr-eom-dbi-m2p2}
\end{equation}
With the quadratic potential $\tilde{V} = A \varphi^{2}$ and the warp factor in form of $W(\varphi)$, the Hubble rate in Eq.~(\ref{dbi-vphi-H}) and the slow-roll equation in Eq.~(\ref{sr-eom-dbi-m2p2}) for scalar field $\varphi$ simplifies to
\ba
3H^2 &=& \frac{1}{\ls\cs\varphi^4} y  \label{dbi-H-y} \,,\\
3H\dot\varphi &\simeq& \frac{2}{\ls\varphi} (1+y)  \label{eom-dbi-m2p2-y} \,,\\
y &=& 1 + \alpha \cs \varphi^2 \label{dbi-m2p2-y}  \,.
\ea
We note that $y>1$ since $\alpha>0$ always holds, so we can immediately obtain the parameters $\delta$ and $\epsilon$ as
\ba
\delta &=& \frac{\dot\varphi}{H\varphi} = \frac{2}{\alpha} {\lrb\frac{y+1}{y}\rrb} (y-1) \label{sr-dbi-m2p2-delta} \,,\\
\epsilon &=& \frac{\dot\varphi^2}{2} \frac{1}{H^2} \frac{1}{\cs\varphi^4} = \frac{2}{\alpha} {\lrb \frac{y+1}{y} \rrb}^2 (y-1) \label{sr-dbi-m2p2-eps} \,.
\ea

We wish to find the relation between the field $\varphi$ and the e-folds $N$ via the relation
\be
N = -\int H dt = -\int \frac{d\varphi}{\varphi} \frac{1}{\frac{\dot\varphi}{H\varphi}} = -\int \frac{d\varphi}{\varphi} \frac{1}{\delta} \label{dbi-efold-vphi} \,.
\ee
According to Eq.~(\ref{dbi-m2p2-y}) we can derive the relation for $dy$ and $d\varphi$ as,
\begin{equation}\label{sr-dbi-m2p2-dy-deps}
\frac{dy}{y-1} = 2\frac{d\varphi}{\varphi} {\lrb 1 + \frac{1}{2} \frac{s}{\delta} \rrb} \sim  2\frac{d\varphi}{\varphi} \,,
\end{equation}
where $s$ is the small parameter defined in Eq.~(\ref{mod-pms}). To approximate the last term in Eq.~(\ref{sr-dbi-m2p2-dy-deps}), we have used the relation\footnote{The approximation $s\ll\epsilon$ cannot tell us $\css\sim1$ even though $s\sim0$. However, in Appendix \ref{dbi-sr-quad-sra} we show that we indeed have $\css\sim O(1)$ if using the relation (\ref{sr-dbi-m2p2-dy-deps}) for approximation.}
\be
\frac{s}{2\delta} = \frac{s}{\epsilon} \frac{y+1}{2y} < \frac{s}{\epsilon} \ll 1 \label{sr-dbi-appro} \,.
\ee
Now we can obtain $N$ under this approximation. Substituting Eq.~(\ref{sr-dbi-m2p2-delta}) into Eq.~(\ref{dbi-efold-vphi}) and applying Eq.~(\ref{sr-dbi-m2p2-dy-deps}),  we have
\be
N = -\frac{\alpha}{4}\int \frac{ydy}{(y-1)^2(y+1)} \,,
\ee
and its solution is,
\be
\frac{8N}{\alpha} = \frac{1}{y-1} + \frac{1}{2}\ln\frac{y+1}{y-1}  \spc\quad (y>1, \cs\simeq1) \label{sr-dbi-m2p2-N-y}\,. 
\ee
For the quadratic potential $V\propto \phi^2$ we will recover a relation for $\varphi^{-2}= \phi^2 = 4N = 2mN$ when $y\gg1$ according to the above equation. This is the slow-roll prediction for the quadratic potential in canonical inflation, where the sound speed is exactly $\css=1$, and hence applies if DBI inflation with a quadratic potential is approximated by canonical inflation. 

Equation~(\ref{sr-dbi-m2p2-N-y}) is the general relation between the scalar field $\varphi$ (or $\phi$) and $N$. Note that $y$ can be either of order 1 or much greater, according to the calculation in Appendix \ref{dbi-sr-quad-sra}, even when we impose the constraint $\css \sim O(1)$. 

We can write the Eq.~(\ref{sr-dbi-m2p2-N-y}) in terms of the Lambert $\mcl W$ function as follows:
\be
y = 1+\frac{2}{\theta(x)} = 1 + \frac{2}{\mcl W(e^{x + 1}) - 1} ,\quad\quad\mbox{(at $\cs\simeq 1$)} \label{sr-dbi-alpha-N} \,,
\ee
where $x={16N}/{\alpha}$. Therefore we can write the $\delta$ and $\epsilon$ according to Eqs.~(\ref{sr-dbi-m2p2-delta}) and (\ref{sr-dbi-m2p2-eps})
\ba
\delta &=& \frac{8}{\alpha} \frac{\mcl W}{\mcl W^2-1} \label{sr-dbi-m2p2-delta-N-theta}\,,\\
\epsilon &=& \frac{16}{\alpha} \frac{\mcl W}{\mcl W^2-1} \frac{\mcl W}{\mcl W+1} \label{sr-dbi-m2p2-eps-N-theta}\,, \\
N &=& \frac{\alpha}{16} (1+\mcl W+\ln{\mcl W})  \label{sr-dbi-m2p2-N-theta} \,. 
\ea
They imply $\epsilon \sim 1/N$. We now can treat Eqs.~(\ref{sr-dbi-m2p2-eps-N-theta}) and (\ref{sr-dbi-m2p2-N-theta}) as parametric equations for $\delta$, $\epsilon$ and $1/N$, respectively, with parameter $\theta$.

By denoting $\beta = \left(1+ 1/{\mcl W} \right)^2$ we can also write
\begin{numcases}{\beta=}
1 \spc & \; $\mcl W \gg 1 \Longleftrightarrow \alpha \ll 16N$ \,,\\
4 \spc &\; $\mcl W \raw 1 \Longleftrightarrow \alpha \gg 16N$ \,.
\end{numcases}
Since this quantity (ratio) is monotonically increasing to 4 with $\alpha$,  it will never contribute a term like $1/N$, so we can regard this ratio $\beta$ as a `pseudo-constant'. Therefore according to Eqs.~(\ref{sr-dbi-m2p2-eps-N-theta}) and (\ref{sr-dbi-m2p2-N-theta}) we represent the parameter $\theta$ as
\be
\theta = \beta \frac{16}{\alpha\epsilon} \quad,\quad \theta\alpha \sim 16\beta{N} \label{sr-dbi-m2p2-theta} \,,
\ee
where the second relation is approximated by setting $\epsilon \sim 1/N$ for later convenience in discussion. The parameters $\theta$ and $\alpha$, due to the relation in Eq.~(\ref{sr-dbi-m2p2-theta}), will be treated interchangeably when we discuss the parameter constraints in a later section. 

We will use these relations to generalise our conclusions below. So far we have completed the slow-roll calculation for DBI inflation with a quadratic potential. Now we need to evaluate the power spectrum $\Ps$ and its spectral index. According to Eq.~(\ref{mod-ps}) and the variables which have been derived in Eqs.~(\ref{dbi-H-y}), (\ref{dbi-m2p2-y}), (\ref{sr-dbi-m2p2-delta-N-theta}), (\ref{sr-dbi-m2p2-eps-N-theta}) and (\ref{sr-dbi-m2p2-theta}), the power spectrum $\Ps$ is written as,
\be
\Ps  = \frac{1}{48\pi^2}\frac{1}{\epsilon} \frac{\alpha^2}{\ls} \theta \left(1+\frac{\theta}{2}\right)  = \frac{1}{96\pi^2} \frac{\alpha^3}{\ls} \frac{(\mcl W^2-1)^2(\mcl W+1)}{W^2} \label{sr-dbi-m2p2-ps-N}  \,,
\ee
where $N_*$ is the e-folds number at the end of the DBI inflation.  According to Eq.~(\ref{sr-dbi-m2p2-theta}), the power spectrum $\Ps$ in Eq.~(\ref{sr-dbi-m2p2-ps-N}) provided from the quadratic potential in DBI inflation can also be rewritten in terms of $\epsilon$ together with model parameters,
\be
\Ps = \frac{1}{3\pi^2} {\lrb  \beta \frac{A}{\epsilon^2} + 8\beta^2 \frac{1}{\ls\epsilon^3} \rrb} \label{sr-dbi-m2p2-ps-eps-2}\,.
\ee
We can see that the slow-roll prediction for the power spectrum has terms due to the scalar potential denoted by its scale $A$, and the warp geometry denoted by the strength $\ls$. This is a new output from our slow-roll calculation, in comparison to the conventional approach where the first term appears only. Since in that case the model has been always studied at $\cs \sim 1$ initially, the DBI action is reduced to the canonical type, which no doubt will only present a limited prediction that is the first term in our generalised equation in Eq.~(\ref{sr-dbi-m2p2-ps-eps-2}).

According to Eqs.~(\ref{mod-ns}) and (\ref{sr-dbi-m2p2-ps-N}), we obtain the spectral index as
\be
\ns = - (2\epsilon + \eta)  \label{sr-dbi-m2p2-ns-N} \,.
\ee
The spectral index here is exactly derived from Eq.~(\ref{sr-dbi-m2p2-ps-N}). There is no significant contribution from the derivatives of the sound speed, since $\cs \sim 1$ or from Eq.~(\ref{sr-dbi-appro}) which tells us that the third parameter $s$ is much less than $\epsilon$. We only need to work out the second parameter $\eta$ via the result for $\epsilon$, so that,
\be
\eta = \frac{\mcl W^2 - \mcl W + 2}{\mcl W^2} \epsilon \label{sr-dbi-m2p2-eta-theta}\,.
\ee
Then by means of Eq.~(\ref{mod-ns}), we have the result for the spectral index,
\be
\ns = -3\epsilon  \frac{\mcl W^2 - \mcl W/3 + 2/3}{\mcl W^2},  \quad (\mcl W>1)\label{sr-dbi-m2p2-ns-eps-theta}\,.
\ee
However, this formula does not give a clear understanding of the spectral index, unlike the one in Eq.~(\ref{sr-dbi-m2p2-ns-N}). However, we can check that the relations for the spectral index will have the same asymptotic behaviour (at $\cs \sim 1$)
\begin{numcases}{\ns=}
-\frac{2}{N_*} & \;  $\mcl W\raw1 \Longleftrightarrow \alpha\gg16N$ \,, \\
-\frac{3}{N_*} & \;  $\mcl W \gg 1  \Longleftrightarrow \alpha\ll16N$ \,,
\end{numcases}
where one can derive the $\epsilon$ as 
\begin{numcases}{\epsilon=}
\frac{1}{2N_*} & \;  $\mcl W\raw1 \Longleftrightarrow \alpha\gg16N$ \,, \\
\frac{1}{N_*} & \;  $\mcl W \gg 1 \Longleftrightarrow \alpha\ll16N$ \,.
\end{numcases}
Finally, in terms of $(N,x;\mcl W(e^{1+x}))$, we present both the spectral index $n_s$ and the tensor-to-scalar ratio $r$ as,
\ba
n_s &=& 1 - \frac{1}{N} \frac{x}{\mcl W-1}\frac{3\mcl W^2-\mcl W+2}{(\mcl W+1)^2}  ,\quad (\mcl W>1) \label{sr-dbi-m2p2-r-W} \,,\\
r &=& \frac{16\cs}{N} \frac{x}{\mcl W-1} \frac{\mcl W^2}{(\mcl W+1)^2}   ,\quad (\mcl W>1, \cs\sim1) \label{sr-dbi-m2p2-r-W2}\,,
\ea
where $x=16N/\alpha$, and the sound speed can be obtained from $\css = 1 - 2\epsilon/[3(\mcl W-1)]$. Since $x>0$ and $\mcl W>1$, the spectral index is always red tilted. We can see that the prediction for both $n_s$ and $r$ are model parameter dependent, in view of $\alpha=M^2\ls/2$. DBI inflation, however, does not provide a simple form for the spectral index $n_s$ and the tensor-to-scalar ratio $r$ as in previous models, where these quantities can be expressed as a function of the variable set $(n,m,N)$. This model has another variable $\alpha$ which can play an important role in determining the value of $n_s$ and $r$. 

These results, for the power spectrum in Eq.~(\ref{sr-dbi-m2p2-ps-eps-2}) and spectral index in Eq.~(\ref{sr-dbi-m2p2-r-W}), are derived in the non-relativistic limit $\cs \sim 1$. Meanwhile, we can evaluate the following relation
\be
\frac{23}{24} < \frac{\mcl W^2 - \mcl W/3 + 2/3}{\mcl W^2} < \frac{4}{3}  \qquad (\mathcal W > 1)\,,
\ee
which appeared in Eq.~(\ref{sr-dbi-m2p2-ns-eps-theta}). Therefore we will have a leading contribution for the spectral index $\ns = O(\epsilon)$ when the DBI inflaton field moves with $\cs\sim1$. 

\bfg[t]
\centering
\includegraphics[width=0.9\textwidth]{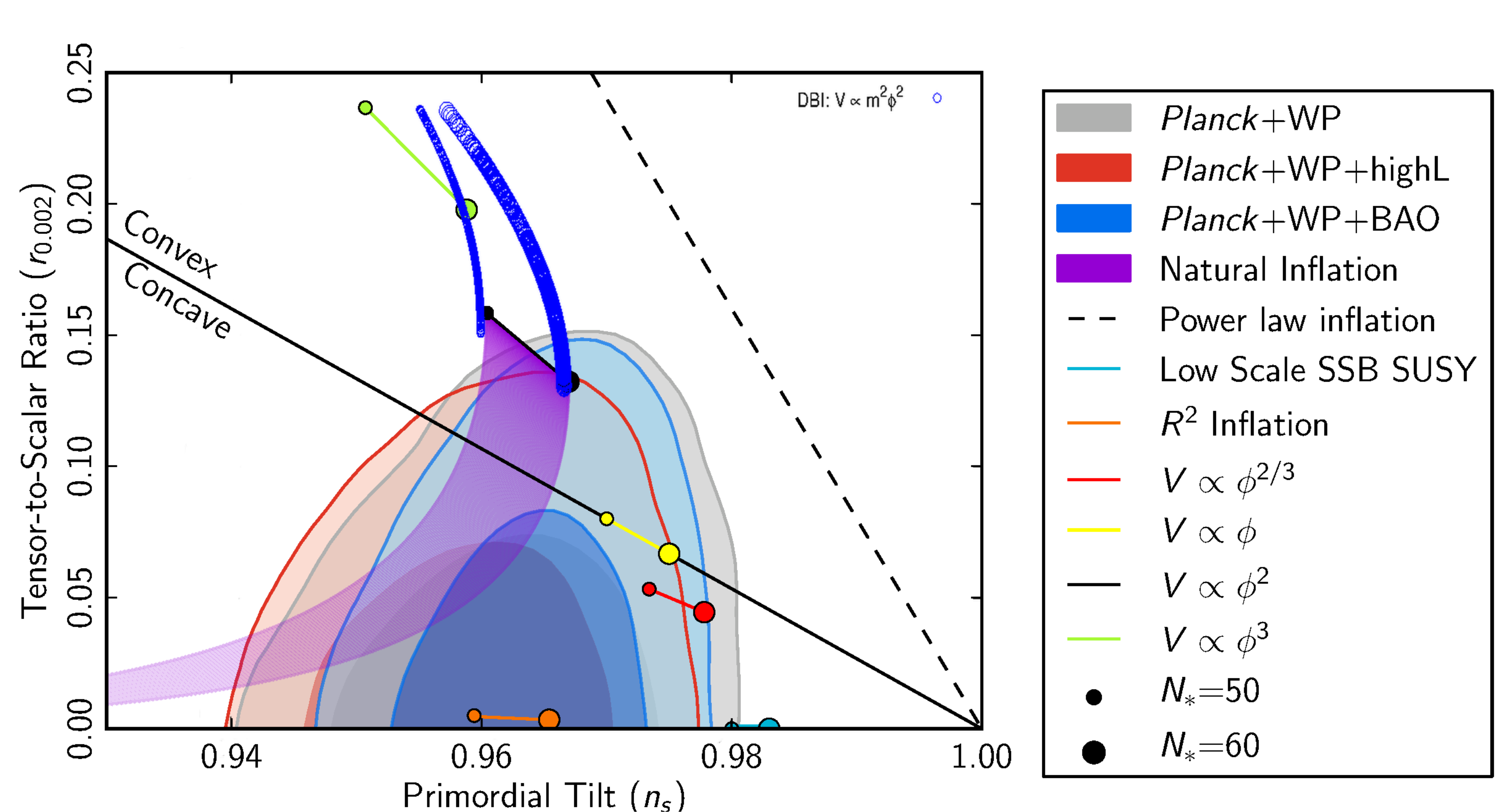}
\caption{Slow-roll predictions for the DBI inflation with quadratic potential and the {\it Planck} constraints on inflation models. [Based on an image from Ref.~\cite{Ade:2013uln}, original image credit ESA/{\it Planck} Collaboration.]}
\label{fig:sr-dbi-m2p2-nsr}
\efg

Figure~\ul{\ref{fig:sr-dbi-m2p2-nsr}} shows the predictions against data. We see that for any available model parameter $\alpha$ DBI inflation has difficulties in providing the required observables, as the results lie above the $m=2$ canonical model which is coming under observational pressure, particularly with $N_* = 50$. But for larger pivot e-folds, for example $N_*=60$, the DBI model with this quadratic potential has the possibility to match observations. Note incidentally that our predictions meet those of natural inflation (the purple shaded region), which approach the canonical quadratic case from below.

Our result is different from the result that $\ns = O(\epsilon^2)$ in Ref.~\cite{Alishahiha:2004eh}. The difference between these results is that ours is in the $\cs \rightarrow 1$ limit, while theirs applies in the warp-factor dominated regime of $\cs \rightarrow 0$, a limit in which we are able to reproduce their result.  To show this we start with the the Hubble parameter $H^2$ and the sound speed $\css$ given by
\be
3H^2 =  \frac{1}{\varphi^4} \frac{y}{\ls\cs} \quad\spc\quad \epsilon = \frac{3}{2} \frac{1-\css}{y} \,.
\ee
So at $\cs\raw0$, which also means $y\gg1$ if we focus on $\epsilon\ll1$, according to Eq.~(\ref{sr-dbi-m2p2-eps}) we have the following approximate relations
\be
3H^2 \simeq \frac{3}{2\ls\epsilon} \frac{1}{\cs\varphi^4}\quad\spc\quad\epsilon \simeq \frac{3}{2} \frac{1}{y} \quad\spc\quad \alpha = \frac{3}{\epsilon^2} \label{ecs0} \,.
\ee
Therefore, according to Eqs.~(\ref{mod-ps}) and (\ref{mod-ns}) we have
\ba
\Ps &=&  \frac{1}{16\pi^2} \frac{1}{\ls\epsilon^2} \frac{1}{\css\varphi^4} = \frac{1}{16\pi^2} \frac{1}{\ls\epsilon^2} \frac{\alpha^2}{(y-1)^2} \label{ps-0}\,,\\
\ns &=& -\left(-2\frac{\epsilon^\prime}{\epsilon} -2\frac{y^\prime}{y-1}\right) = 2\frac{y^\prime}{y} \frac{1}{y-1} \label{ns-0}\,.
\ea
Using the relation in Eq.~(\ref{ecs0}), we can simplify them to be
\ba
\Ps &=& \frac{1}{4\pi^2}\frac{1}{\ls\epsilon^4} \,,\\
\ns &=& \frac{4}{3}\epsilon\eta \propto O(\epsilon^2)\,,
\ea
since $\epsilon,\, \eta=-\epsilon^\prime/\epsilon=y^\prime/y$ are of the same order. As this regime predicts that $\ns$ is very close to one, it is under strong pressure from observations.

To summarise our results in the slow-roll approximation, we obtained formulae for the power spectrum $\Ps$ in Eq.~(\ref{sr-dbi-m2p2-ps-N}) or (\ref{sr-dbi-m2p2-ps-eps-2}) and its spectral index $n_{{\rm s}}$ in Eq.~(\ref{sr-dbi-m2p2-ns-N}). We present all the relations in each limit in Table \ul{\ref{sr-dbi-m2p2-obs-theta}}. 

\btb[t]
\centering
\def\arraystretch{1.}
\bt{|c|c|c|c|c|c|c|c|}
\hline
$\theta$ or $\alpha$ & $\frac{1}{N_*}$& $\epsilon(\theta)$ &$\epsilon(N_*)$ & $\Ps(\epsilon)$ & $\Ps(N_*)$ & $\ns$ & Domination \\
\hline
$\theta \raw 0$ or $\alpha \gg 16N_*$ & $\frac{8}{\alpha\theta}$ & $\frac{4}{\alpha\theta}$& $\frac{1}{2N_*}$ & $ \frac{A}{\epsilon^2}$ & $\propto AN_*^2$& $-\frac{4}{2N_*}$ & Scalar potential \\
$\theta \gg 1$ or $\alpha \ll 16N_*$ & $\frac{16}{\alpha\theta}$ & $\frac{16}{\alpha\theta}$& $\frac{1}{N_*}$ & $ \frac{1}{\ls\epsilon^3}$ & $\propto \frac{N_*^3}{\ls}$ & $-\frac{6}{2N_*}$ & Warp geometry \\
\hline
\et\caption{Predictions for the observables in DBI inflation with a quadratic potential, in each limit for $\theta$. At other values of $\theta$, the outcome is determined by both the scalar potential and the warp geometry (see Eq.~(\ref{sr-dbi-m2p2-ps-N}) or Eq.~(\ref{sr-dbi-m2p2-ps-eps-2})). These results are obtained under the assumption in Eq.~(\ref{sr-dbi-m2p2-dy-deps}), which requires the sound speed $\css \sim O(1)$. Also note that $\theta = \mcl W - 1$.}\label{sr-dbi-m2p2-obs-theta}
\etb

Finally, it is worth mentioning that we can also make predictions for the case where the sound speed $\css \ll 1$ in the same manner that we have just applied. We can obtain the solution in this relativistic case, but the relation $\epsilon=\epsilon(N)$ is not obviously applicable from the e-folds $2/(y^2-1)\propto\mathcal{W}(N)$ which, according to Eq.~(\ref{dbi-z-css-0}), can be obtained from Eq.~(\ref{dbi-efold-vphi}). These models are subject to the non-gaussianity constraint that the sound speed be greater than about 0.07 \cite{Ade:2013ng}.

\sect{Model Parameter Estimation}\label{sect-mpe}

We now discuss parameter estimation for the models investigated in the previous sections. Throughout this paper, the potential considered possesses only one free parameter. Since the amplitude of the scalar power spectrum is accurately determined to be about $2.5\times 10^{-9}$ from observational data, we can now estimate the model parameters for each case, for example taking the pivot e-folds to be $N_{*}=50$.  

\subs{Sum-separable models}
We take the expression of the scalar power spectrum from Eq.~(\ref{ps-ssm}) and apply a (base 10) logarithm, giving 
\ba
\log{A} &=& \frac{m}{2n} \left[  - \frac{6.53 + (1+\frac{m}{\beta})\log{N} + \log(\frac{\beta}{m}\sqrt{2n-1})}{\frac{m}{\beta}\frac{1}{2n-1}} - \log \left(\frac{m6^{n-1}}{n}  \beta^{2n-1} \right) \right] \,.\label{ssm-mpe}
\ea
Note that the model parameter $A$ denotes the amplitude appearing in the potential $V(\phi) = A\phi^{m} = \lambda_p\phi^{m}/m$ after the coefficient of the kinetic term has been rescaled to unity. For example, for  the quadratic potential $A = M^{2}/2$ where $M^{2}$ is the mass of field, while for  the quartic potential $A = \lambda/4$.

As an application of this formula, we consider the canonical inflation cases where $n=1$ and $m=2,4$. Then we will obtain an estimate for the mass scale for the quadratic and quartic potential respectively. Then
\ba
\log M^2 &\simeq& -10.2 \,,\\
\log\lambda &\simeq& -12.5 \,.
\ea
For non-canonical inflation models, a similar analysis can be performed.

\subs{Tachyon models}
For the Tachyon model, we consider the results in Section~\ref{app-to-sect-tachyon}. From Eq.~(\ref{tachyon-ps-N}) we find the relation between $A$ and $\ls$ to be
\ba
\log\ls &=& \frac{2}{m}\log{A} + 2(\mbf{A} + \mbf{B} + \mbf{C}) + {(1+2\mu)\log{(2N)}} \\
\mbf A &=& -\mu \log{(12\pi^{2} \times 2.5\times10^{-9})} = 6.53\mu \nonumber ,\,
\mbf B = \log{m}+ \left(\frac{1}{2}+\mu\right)\log{\mu} ,\,
\mbf C = -\frac{1}{m}\log{\cs} \,.
\ea
The equation for $\mbf{C} \sim 0$ above is approximate due to the condition $1/3< \cs < 1$, and particularly $\cs \sim 1$ if $\epsilon \ll 1$. Therefore, for the quadratic potential $A = M^{2}/2\,, m=2 \,, \mu = 1$, we can evaluate the parameters as,
\be\label{tachyon-num-m2p2}
\log\ls = \log{\frac{M^{2}}{2}} + 19 + 2\log{2} \simeq \log{M^{2}} + 19.3 \,.
\ee
For the quartic potential, $A = \lambda/4 \,, m=4 \,, \mu=3/4$, the parameters are
\be\label{tachyon-num-lp4}
\log\ls = \frac{1}{2}\log{\frac{\lambda}{4}} + 18.2 \simeq \frac{1}{2}\log{\lambda} + 17.9 \,.
\ee
We have carried out a Monte Carlo Markov chain fit to the data through an extension of our analysis in Ref.~\cite{Li:2012vt} and found results which are in agreement, shown in Figure~\ref{f:tachnum}.  Similar results can be found in Ref.~\cite{Devi:2011qm}.

\bfg[t]
\centering
\includegraphics[width=0.4\textwidth]{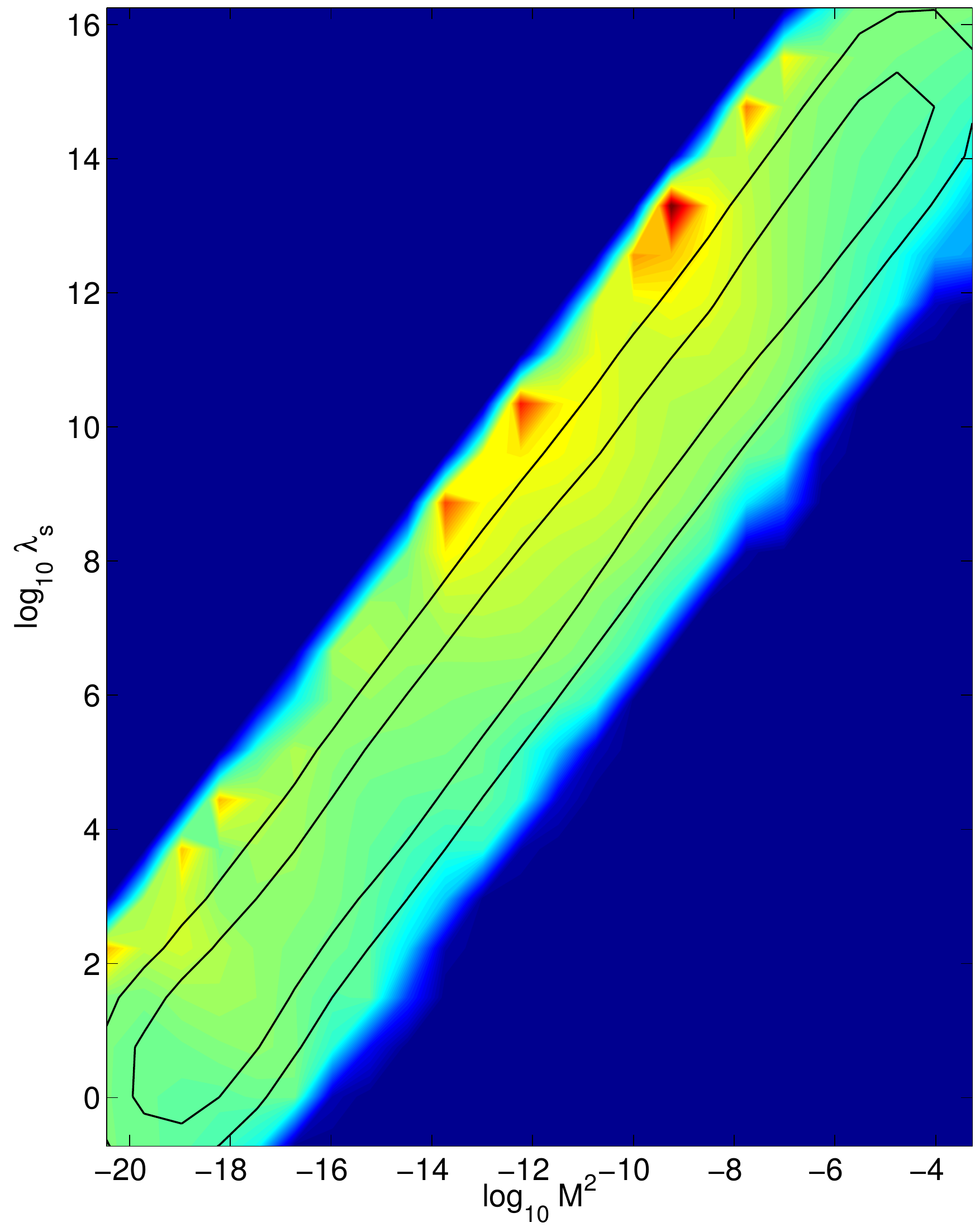} \hspace*{1cm}
\includegraphics[width=0.4\textwidth]{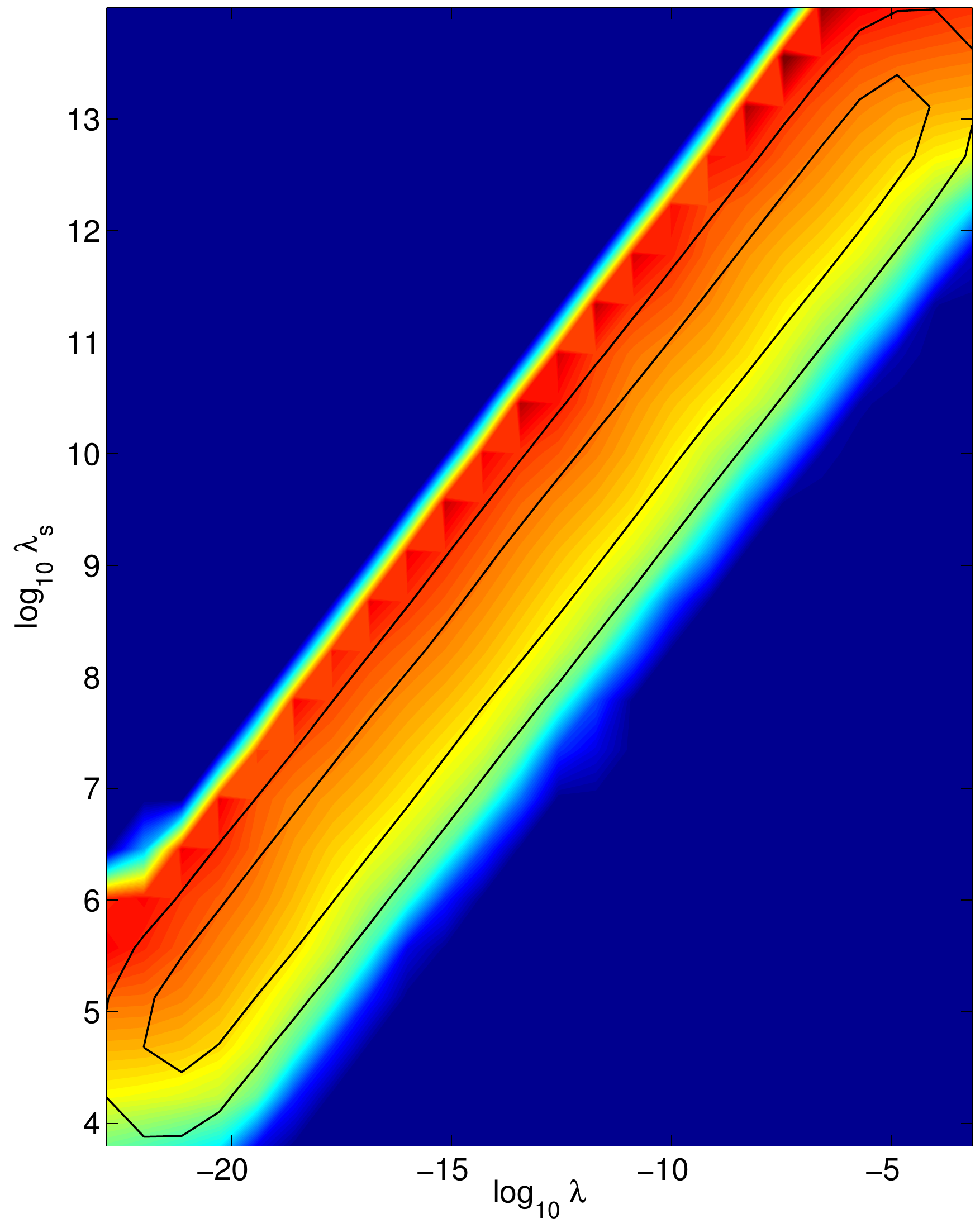} 
\caption{MCMC constraints on model parameters for the quadratic (left) and quartic (right) tachyon cases. The exact degeneracy between the potential normalisation and the tachyon parameter predicted by our analytic analysis is clearly seen.}
\label{f:tachnum}
\efg

\subs{DBI models}
\ssubs{Quadratic potential}
We only study the parameter estimation for the case of $\css \sim 1$, as in Section~\ref{sect-dbi-m2p2-sra}, as otherwise that it is hard to find the required function $\epsilon = \epsilon(N)$. For this reason, for the quadratic potential we present the results in the {\it non-relativistic case} only.

According to the results in Section \ref{sect-dbi-m2p2-sra}, though the formulae for the power spectrum are complicated, we can still approximate the value for $\alpha$ which encodes the scale of the potential ($A$) and the strength of the warped geometry ($\ls$). According to Eq.~(\ref{sr-dbi-m2p2-ps-N}), the combined contribution from these indicates that
\be\label{sr-dbi-m2p2-mpe-aes}
\theta \sim 2 \quad{{\rm or}}\quad \alpha \gtrsim 8N \,.
\ee
Taking the logarithm of this relation (\ref{sr-dbi-m2p2-mpe-aes}) at the pivot scale $N_*=50$ we have the linear equation
\be
l \equiv \log M^2 + \log\ls \simeq 2.9 \label{sr-dbi-log-alpha-N} \,.
\ee
The relation in Eq.~(\ref{sr-dbi-m2p2-mpe-aes}) gives us two possibilities for the locations or choices for the model parameters $A \,, \ls$ at some critical value such as $\alpha \sim 8N_*$. We will see according to Eq.~(\ref{sr-dbi-log-alpha-N}) and Eq.~(\ref{dbi-vphi-H}) that larger $\alpha$ implies a strong contribution from the scalar field potential. In the other case, $\alpha$ smaller than the critical $8N_*$, inflation will be dominated by the warped geometry.

Recalling Table~\ul{\ref{sr-dbi-m2p2-obs-theta}}, we already have the relation between $\theta$ and e-folds $N$. Therefore we can approximate the model parameter range, as presented in Table~\ul{\ref{sr-dbi-m2p2-mod-pms-bound}}. This table also indicates the value $\log m^2 + \log \ls \sim 3$ (or similar) should be found in parameter space. Above this critical value DBI inflation will be dominated by the scalar field potential, while on the other side the warp geometry will dominate. 

\btb[t]
\centering
\def\arraystretch{1.2}
\bt{|c|c|c|}
\hline
	$\alpha$ & $\log M^2 + \log \ls$ & Domination\\
\hline
$\alpha \gg 8N_*$ & $>2.9$  & Scalar potential\\
$\alpha \ll 8N_*$ & $<2.9$ & Warp geometry \\
\hline
\et\caption{Constraints on the model parameters in DBI inflation with quadratic potential. The model parameters, $A$ and $\ls$, are correlated along the line $l \defeq \log M^2 + \log \ls$. The parameter $\alpha = A\ls = M^2\ls/2$, where $M^2$ is the mass scale of the scalar potential and $\ls$ the strength of the warp factor.}\label{sr-dbi-m2p2-mod-pms-bound}
\etb

\ssubs{Quartic potential}
The power spectrum was obtained in Section \ref{sect-eval-dbi-lp4}. Following the discussion there, we can approximate the prediction for model parameters at the $N_* = 50$. We note that in this case, although we can still constrain model parameters by requiring the correct amplitude of density perturbations at the pivot scale, the 
models typically do not lie in the within the 95\% confidence region of the $n_s$--$r$ plane.
\bi
\item[$\star$] {\it Non-relativistic case}\\
According to Eq.~(\ref{dbi-sr-ps}) in the non-relativistic limit where $\cs \sim 1$
\be\label{kmc-dbi-sr-m2p2-mpe-nonrel}
\log \ls - \log(1+\alpha) \simeq 13.1 \,,
\ee
for the parameter correlation. We can estimate the bound for each parameter with different $\alpha$. If $\alpha \ll 1$ then we can expect the low bound value for the strength $\ls$ for warp factor,
\be
\log \ls \gtrsim 13.2 \label{sr-dbi-lp4-nrel-lambda-1} \,,
\ee
while in the limit $\alpha \gg 1$, we will have the upper bound for strength $\lambda$ of the scalar potential,
\be
\log\lambda < - 13.1 + \log4 \simeq -12.5 \label{sr-dbi-lp4-nrel-lambda-2} \,.
\ee
This is the observational value for canonical inflation with a quartic potential.

\item[$\star$] {\it Relativistic case}\\
Similarly we consider the case $\cs \ll 1$ from Eq.~(\ref{dbi-rel-Ps-N}), obtaining
\be\label{kmc-dbi-sr-lp4-mpe-rel}
\log\ls \simeq  13.8\,.
\ee
Hence to have relativistic motion during inflation, the strength of the warp geometry must take this value. We cannot see a relation for the parameter $A$ for the potential if we just consider the power spectrum in Eq.~(\ref{dbi-rel-Ps-N}), but if we recall the footnote in Section \ref{sect-eval-dbi-lp4} when deriving  Eq.~(\ref{dbi-rel-Ps-N}), we have an asymptotic condition relevant to both parameters $A$ and $\ls$:
\be
\alpha \epsilon \cs \sim 1.5 \,,
\ee
where $\epsilon = 2/N$ in the relativistic case. So the asymptotic relation between $\lambda$ and $\ls$ is
\be\label{kmc-dbi-sr-lp4-mpe-rel-2}
\log \ls + \log \lambda \sim \log\frac{3N_*}{\cs} \simeq  2.2 - \log\cs \,.
\ee
We can evaluate this equation at some values of $\cs$. For example, for $\cs \sim 0.01$, which is of the same order as $\epsilon$, we will have $\log \ls + \log \lambda \sim 4.2$.
\ei
All these parameter estimation cases have also been examined using MCMC methods, and the results from our CosmoMC \cite{Lewis:2002ah} exploration are presented in our companion paper II.

\sect{Conclusions}\label{sect-conclusion}

In this paper we have developed and applied a systematic method for deriving observational predictions in non-canonical single-field inflation models, using the slow-roll approximation encoded as a differential equation for $u = 1/\epsilon$ rather than as a set of slow-roll parameters. We have given explicit calculations for several such models, including the Tachyon and DBI cases, deriving observables such as the power spectrum $\Ps$ and its spectral index  $n_s$ in terms of e-folds $N$.  For some models we also present the results in terms of slow-varying parameters rather than $N$, when we are unable to explicitly solve the transcendental equation for $\epsilon$ and $\cs$, for example in DBI inflation with quartic potential. 

The use of field redefinition is another key methodology in this paper. It can simplify the process of finding the solution for $u$, and in terms of the redefined scalar field the reshaped potential can reveal the degeneracy of model parameters which includes both the strength of the kinetic energy and the strength of the potential energy. By this method, in Tachyon models we have obtained an explicit correlation of $f$ and $\lambda$ which has previously been found only via numerical calculation. For the DBI inflation models, we have also obtained for the first time a similar formula describing the correlations between its two model parameters, though they are not as explicit as the relation in the Tachyon  model.

While this method cannot give a full exact solution, as obtained either via non-slow-roll approaches or by numerical exploration, it can nevertheless offer some advantages for modelling inflationary cosmology. For one thing it may suggest to reconstruct the Lagrangian which will potentially give an explicitly solvable relation for $\css=\css(u)$. For another, it can provide a quick parameter estimation since the power spectrum, which is accurately determined from the observational data, is also formulated via slow-varying parameters. In other words, given the Lagrangian or potential, by finding the $p_{,X\varphi}$ and then the variable $u$, we can solve the power spectrum by Eq.~(\ref{mod-ps}) which constrains the model parameters for the considered model. For some classes of inflation model, this process will be quite straightforward, such as the sum-separable class where $p_{,X\varphi}=0$ and the product-separable class where $p_{,X\varphi}\varphi^{\prime}/p_{,X}=-2/u$.

For DBI inflation models with different potentials, we adopted different approaches due to their individual complexities. The DBI inflation with a quadratic potential, if not immediately using limiting cases of $\cs$, is the least tractable model in the current paper. However we still found predictions for this case, though we have only presented in detail the non-relativistic predictions for this potential. Additionally, in the relativistic case we have  obtained predictions matching those derived in the conventional manner using the field speed limit  $\cs=0$ \cite{Alishahiha:2004eh}. Due to the irreversible relation for $\epsilon=\epsilon(N)$ from the e-folds integration $2/(y^2-1)=\mcl{W}(N)$ we have not presented the details, but they can readily be computed if one is interested in the application of the method. The case of DBI inflation with a quartic potential is examined in both the non-relativistic and relativistic cases. For both models, not only do we recover the same results as conventional treatments, but our analysis gives a novel formula for the power spectrum, Eq.~(\ref{sr-dbi-m2p2-ps-eps-2}).

The proposed approach is not only able to address monomial potentials, but can also be applied to models with other potentials. For future applications, reconstruction of the inflation model will be an important direction. For the models we have described in this paper, we hope the method will be useful in that more general context. 

A benefit of considering non-canonical Lagrangians is that model predictions can be in better agreement with current data than the canonical case, where even the quadratic potential is starting to come under pressure. We have affirmed previous results \cite{Li:2012vt,Unnikrishnan:2012zu,Unnikrishnan2} showing that the non-canonical $X^n$ models give predictions that, for a given potential, are in better agreement with data. In this article we have extended that conclusion to the Tachyon models, most explicitly via the field redefinition approach which shows that under slow-roll they are equivalent to canonical models with reshaped potentials of shallower slope, as favoured by the data. For DBI models the situation is less promising; with a quadratic potential the predictions in the non-relativistic limit, seen in Fig.~\ref{fig:sr-dbi-m2p2-nsr}, are further from scale-invariance than the canonical case, while in the relativistic limit the cancellation of the leading-order slow-roll correction noted in Ref.~\cite{Alishahiha:2004eh} places the model too close to scale-invariance. In the quartic case the situation is even less promising. We conclude therefore that DBI models with simple potentials are in significant tension with current data.

\acknowledgments
S.L.\ was supported by a Sussex International Research Scholarship, and A.R.L.\ by the Science and Technology Facilities Council [grant numbers ST/I000976/1 and ST/K006606/1] and a Royal Society--Wolfson Research Merit Award. We thank Antony Lewis, Hiranya Peiris, and David Seery for useful discussions. 

%
\input{appendix}

\end{document}

%% file: appendix.tex
\appendix

\sect{Validity of the Slow-roll Approximation for DBI inflation}\label{dbi-sra-validity}

Here we present further support for our calculations, notably the slow-roll approximation made in this paper, for DBI models with different scalar potentials $V(\varphi)$. For consistency with the main body of this paper, we will only discuss the redefined scalar potential $\varphi$ that puts the Lagrangian into the form of Eq.~(\ref{e:330}), rather than the original $\phi$.

\subs{Preliminaries}
We carry out the proofs based on the modified equation of motion for DBI scalar field $\varphi$ in Eqs.~(\ref{dbi-phi-vpahi}) to (\ref{dbi-vphi-H}). When considering the predictions under the slow-roll approximation, according to  Eq.~(\ref{dbi-vphi}),
\be
\frac{\ddot\varphi}{\cs^{2}}  + 3H \dot\varphi + \frac{W^{\prime}}{W} \frac{1}{\ls}{\lrb1 + \frac{m}{4}\ls\cs\tilde{V}\rrb} = 0 \,,
\ee
we assumed that the first term $\ddot\varphi/\css$ is negligible compared to the other terms, specifically the second term $3H\dot\varphi$. We start our proof with the following equations,
\def\rbP{{\rm{\bf P}}}
\def\rbQ{{\rm{\bf Q}}}
\def\rbM{{\rm{\bf M}}}
\def\rbN{{\rm{\bf N}}}
\ba \label{dbi-m2p2-sra-eq}
3H^2 = \frac{1}{\ls\cs\varphi^4}\rbP &\spc& \rbP = 1 + \alpha \cs \varphi^{(4-m)} \label{dbi-h2-vld} \,,\\
3H\dot\varphi \simeq \frac{4}{\ls\varphi} \rbQ &\spc& \rbQ = 1 + \frac{m}{4}\alpha \cs \varphi^{(4-m)} = 1+\frac{m}{4}(\rbP-1)\label{dbi-sra-eom-app-vld} \,,\\
\css = 1-\ls{\dot\varphi}^2  &\spc&  \alpha = A\ls \label{dbi-css-vld} \,, 
%
\ea
where the potential $\tV=A\varphi^{4-m}$ and warp geometry $W=1/\varphi^4$ are taken to satisfy the simplified equations above. 

\subs{Evaluation of the validity of the slow-roll approximation}

We define a quantity $\rm E$ as
\be
{\rm E} = \frac{\ddot\varphi/\css}{3H\dot\varphi} \label{dbi-sra-defs-E}\,.
\ee
We aim to find a relation between $\rm{E}$ and a limited number, optimally one, of the small parameters such as $\epsilon$ defined above in Eq.~(\ref{dbi-sra-eps-validity}), since it can be constrained to $\epsilon \in (0, 1)$ by observational data, and to prove whether the slow-roll approximation is sufficiently good for DBI inflation by evaluating whether $\rm{E}/\epsilon$ is sufficiently small.

We need to work out some auxiliary parameters according to Eqs.~(\ref{dbi-h2-vld}), (\ref{dbi-sra-eom-app-vld}), and (\ref{dbi-css-vld}) as follows:
\ba
\hat{\rbP} &=& \frac{\dot{\rbP}}{H\rbP} = \Big(s+(4-m)\delta\Big) \frac{\rbP-1}{\rbP} \label{dbi-sra-hatp-validity}\,,\\
\hat{\rbQ} &=& \frac{\dot{\rbQ}}{H\rbQ} = \Big(s+(4-m)\delta\Big) \frac{\rbQ-1}{\rbQ}\label{dbi-sra-hatq-validity} \,,\\
\delta &=& \frac{\dot\varphi}{H\varphi} = 4\cs\varphi^2 \frac{\rbQ}{\rbP} \label{dbi-sra-delta-validity} \,,\\
\epsilon &=& -\frac{\dot H}{H^2} = -\frac{1}{2} {\lrb \hat{\rbP} - s - 4\delta \rrb} \label{dbi-sra-eps-validity}\,,\\
\tilde\eta &=& \frac{\ddot\varphi}{H\dot\varphi} = \hat{\rbQ} + \epsilon - \delta \label{dbi-sra-teta-validity}\,,\\
s &=& \frac{\dot\cs}{H\cs} = -\frac{1-\css}{\css} \frac{\ddot\varphi}{H\dot\varphi} = -\frac{1-\css}{\css}\tilde\eta \label{dbi-sra-s-validity} \,.
\ea
Combining Eqs.~(\ref{dbi-sra-eps-validity}) and (\ref{dbi-sra-teta-validity}), and replacing $\hat{\rbP},\;\hat{\rbQ}$ by Eqs.~(\ref{dbi-sra-hatp-validity}) and (\ref{dbi-sra-hatq-validity}), we obtain
\ba
\tilde\eta &=& 3\delta + s - \epsilon + {\Big( s + (4-m)\delta \Big)} \frac{\rbQ-\rbP}{\rbP\rbQ} 
	= \rbM\delta  - \epsilon + s \Big( 1 + \Lambda \Big) \,,\\
\rbM &=& \Big( 3 + (4-m) \Lambda \Big) \quad\spc\quad \Lambda = \frac{\rbQ-\rbP}{\rbP\rbQ} \label{dbi-sra-teta-M} \,.
\ea
According to Eq.~(\ref{dbi-sra-eps-validity}) we can denote $\delta$ as,
\be
\delta = \frac{2}{\rbN}\epsilon - \frac{1}{\rbP\rbN}s  \quad\spc\quad \rbN =  m + (4-m)\frac{1}{\rbP}   \label{dbi-sra-delta-PN} \,,
\ee
and further by replacing parameter $s$ in terms of $\tilde\eta$ via Eq.~(\ref{dbi-sra-s-validity}) we can reach the formula for $\tilde\eta$ in terms of $\epsilon,\;\css$ and $\rm{\rbP}$,
\ba
\frac{\tilde\eta}{\css} &=&  \frac{2\frac{\rbM}{\rbN}-1}{1 +  (1-\css)(\Lambda - \frac{\rbM}{\rbN}\frac{1}{\rbP})} \epsilon \,.
\ea
Note $\rbN$ is different from the e-folds number $N$. Recalling the definition of $\rm{E}$ we can relate it to the above equation,
\be
\frac{\tilde\eta}{\css} \equiv \frac{\ddot\varphi/\css}{H\dot\varphi} = 3\rm{E} \,,
\ee
and therefore we find it to be
\be
\rm{E} = \frac{\epsilon}{3} \frac{ 2\rbM/\rbN -1 }{1 +  (1-\css)(\Lambda - \rbM/\rbN\rbP)} \label{dbi-sra-E-eps-css-M-N-P-vld}  \,. 
%
\ee
We have a relation between $\epsilon$ and $1-\css$ coming from the definition of $\epsilon$ in Eq.~(\ref{dbi-sra-eps-validity}),
\be
\epsilon = \frac{3}{2} \frac{1-\css}{\rbP} \,.
\ee
Thanks to this we can simplify our prediction for $\rm{E}$ to
\be
\rm{E} = \frac{\epsilon}{3} \frac{ 2\frac{\rbM}{\rbN} -1 }{1 +  \frac{2}{3}(\rbP\Lambda - \frac{\rbM}{\rbN})\epsilon} \,.\label{dbi-sra-E-eps-M-N-PL-vld}
\ee
Instead of $\Lambda$ and $\rbM/\rbN$ we introduce $\Delta = \rbP\Lambda$ as well as $\Gamma = \rbM/\rbN$. We can obtain the new quantities as
\ba
\Delta = \frac{\rbQ-\rbP}{\rbQ} &=& \frac{m-4}{m + \frac{4}{\rbP-1}}  \label{dbi-sra-D-P} \,,\\
\Gamma = \frac{\rbM}{\rbN} &=& \frac{3 + \Delta \frac{(4-m)}{\rbP}  }{m+\frac{(4-m)}{\rbP}}  \label{dbi-sra-G-P}  \,,\\
\Gamma - \Delta &=& \frac{3-m\Delta}{m + \frac{4-m}{\rbP}} \label{dbi-sra-G-D-m} \,,
\ea
and so we can write Eq.~(\ref{dbi-sra-E-eps-M-N-PL-vld}) in a new form,
\ba
\rm{E} &=& \frac{2\epsilon}{3} \frac{ \Gamma - \frac{1}{2} }{1 -  \frac{2}{3}(\Gamma - \Delta)\epsilon} \label{dbi-sra-E-eps-M-N-PD-vld} \,.
\ea
It is still not easy to interpret $\rm{E}$, so we need to simplify this equation. We introduce a useful new transformation defined by
\be
\theta = \frac{2}{\rbP-1} \spc\qquad (\theta>0) \label{dbi-sra-theta-P}\,,
\ee 
and then we write the $\Gamma,\;\Delta$ as,
\ba
\Delta &=& \frac{m-4}{m+2\theta}\,,\\
\Gamma &=& \frac{1}{2}\frac{6\theta^2-(m^2-11m+4)\theta+6m}{(m+2\theta)^2} \,.
\ea
We now introduce two functions,
\def\rbF{{\rm{\bf F}}}
\def\rbG{{\rm{\bf G}}}
\def\rbH{{\rm{\bf H}}}
\ba
\rbF(\theta;m) = \Gamma - \frac{1}{2} &=& \frac{1}{2}\frac{2\theta^2 - (m^2-7m+4)\theta - (m^2-6m)}{(m+2\theta)^2} \label{dbi-sra-defs-F-t} \,,\\
\rbG(\theta;m) = \Gamma - \Delta &=& \frac{1}{2}\frac{6\theta^2-(m^2-7m-12)\theta-2(m^2-7m)}{(m+2\theta)^2} \label{dbi-sra-defs-G-t} \,,
\ea
so that we can simplify Eq.~(\ref{dbi-sra-E-eps-M-N-PD-vld}) further to
\be
\rm{E} = \frac{2\epsilon}{3} \frac{\rbF(\theta;m)}{1 - \frac{2\epsilon}{3}{\rbG(\theta;m)}} \label{dbi-sra-E-F-G-m-t} \,.
\ee

We define a transformation
\be
x = \frac{1}{m+2\theta}  \label{dbi-sra-defs-x}\,,
\ee
which is defined either in the domain $(0,1/m)$ if $m>0$,  or in $(1/m,\infty)$ if $m<0$.\footnote{In this case a divergence will happen when $2\theta = -m$ because of $\theta>0$.} Then we have the following elegant representations
\ba\label{dbi-sra-F-G-x}
\rbF(x;m) &=& \frac{m(m - 4)^2}{4} x^2 - \frac{(m-1)(m-4)}{4}x + \frac{1}{4}   \,,\\
\rbG(x;m) &=& \frac{m(m - 4)^2}{4} x^2 - \frac{(m+3)(m-4)}{4}x  + \frac{3}{4}  \,.
\ea
Therefore we can now reformulate the ratio $\rm{E}$ in Eq.~(\ref{dbi-sra-E-F-G-m-t}) to,
\be
\rm{E} = \frac{2\epsilon}{3} \frac{\rbF(x;m)}{1-\frac{2\epsilon}{3}\rbG(x;m)} \label{dbi-sra-E-F-G-m-x} \,.
\ee

The potential need not necessarily have $m>0$. The relation in Eq.~(\ref{dbi-sra-E-F-G-m-x}) is the function which we sought and is needed for evaluating the validity of the slow-roll approximation in DBI inflation models. Now we want to know whether there is any chance of $\left|\rm{E}\right| \gg 1$. To explore this, we need Eq.~(\ref{dbi-sra-E-F-G-m-x}) and the following inequality, due to the fact that the inequality $\left|\rbF(x;m)\right| \leq \left|\rbG(x;m)\right|$ always holds,
\be
\rm{E}  < \frac{\frac{2\epsilon}{3}\rbG(x;m)}{1-\frac{2\epsilon}{3}\rbG(x;m)} \,.
\ee
Then we will have,
\begin{numcases}{\rm{E}}
\ll 1 \spc & \; $(\left|\frac{2\epsilon}{3}\rbG(x;m)\right| \ll 1)$ \label{dbi-sra-E-vld-bound-1}\,,\\
\simeq -1  \spc & \; $(\left|\frac{2\epsilon}{3}\rbG(x;m)\right| \gg 1)$ \label{dbi-sra-E-vld-bound-2} \,.
\end{numcases}
Further we need to check if there is any singularity, for example $1-(2\epsilon/3)\rbG(x;m) = 0$ in  Eq.~(\ref{dbi-sra-E-F-G-m-x}). We can investigate the boundary values for the function $\rbG(x;m)$ by expressing this function $\rbG(x;m)$ as
\be
\rbG(x;m) = \frac{m}{4} {\lrb (m-4)x - \frac{1}{2}\frac{m+3}{m} \rrb}^2 - \frac{(m-3)^2}{16m} \,.
\ee

Now we investigate in detail under what condition the function $\rm{E}$ will be less than, for example $10 \times\epsilon \lesssim 0.1$. In other words, under what condition will the slow-roll approximation still be valid if we use the approximation in Eq.~(\ref{dbi-sra-eom-app-vld}). The conclusion is determined by the potential we choose for a particular DBI inflation model. So we examine the possibilities in view of a potential with $m>0$ below.
\bi
\item[$\star$] $0<m<4$\\
Considering the potential $V\propto\phi^m$ where $0<m<4$, in turn for the domain $x \in (0, 1/m)$ we will have the relation for the stationary point $x_0 = \frac{1}{m}\frac{m+3}{2(m-4)}$,
\be\label{dbi-sra-vld-xs1-Domain}
x_0 < x_{D^-} < x_{D^+}\,,
\ee
where $x_{D^-}=0$ is the left boundary, while $x_{D^+}=1/m$ is the right boundary. Therefore the lower bound for this function is $\rbG(x_{D^-};m)$. The upper bound is located at $x\raw 1/m$ denoted by $\rbG(1/m;m)$,
\ba
\rbG\Big|_{\rm min} = \rbG(0,m) &\equiv& \frac{3}{4} \,,\\
\rbG\Big|_{\rm max} = \rbG(\frac{1}{m},m) &=& \frac{7}{m} -1 \,.
\ea
As $0<m<4$ we can give the boundary values by the inequalities,
\begin{numcases}
{\rbG\Big|_{\rm max} } 
> 6 \spc& \;  $(0<m<1)$ \label{dbi-sra-E-vld-bound-3} \,,\\
\in (\frac{3}{4}, 6] \spc& \;  $(1\leq m<4)$ \label{dbi-sra-E-vld-bound-4} \,.
\end{numcases}
Equation (\ref{dbi-sra-E-vld-bound-3}) implies that we will have
\be
\rbG(x;m) = \frac{3}{2\epsilon}\,,
\ee
which gives
\be
\quad m = \frac{14\epsilon}{3+2\epsilon} \label{dbi-sra-G-m-eps-invld}\,.
\ee
Assuming $\epsilon \sim 0.01$, this condition tells us the singularity will occur if $m$ is of order $m \ll O(0.05)$, and then the slow-roll approximation will be violated in DBI inflation with a single-term polynomial potential $V=A\phi^m$. In turn, it also suggests that if
\be
\frac{1}{20} < m_c < 4 \,,
\ee
we can still use the slow-roll approximation.

\item[$\star$] $m>4$\\
We can just re-use the relation in Eq.~(\ref{dbi-sra-vld-xs1-Domain}),
\be\label{dbi-sra-vld-xs4-Domain}
x_{D^-} < x_0 < x_{D^+} \,,
\ee
so that the lower bound is $\rbG(x_0;m)$ and the upper bound is $\rbG(x_{D^-};m) \equiv 3/4$. Then we check the lower bound for $\rbG(x_0;m)$,
\be
\rbG(x_0;m) = -\frac{(m-3)^2}{16m} < 0 \,,
\ee
and also we need the relation for both functions $\rbF(x;m),\;\rbG(x;m)$ by using Eqs.~(\ref{dbi-sra-defs-F-t}) and (\ref{dbi-sra-defs-G-t}),
\be
\rbF = \rbG + \Delta - \frac{1}{2} \,,
\ee
is roughly around $\rbG$ since $m>4$ even $m\ll4$. The value of $\left|\rm{E}\right|$ in this case will be in the range of the case Eq.~(\ref{dbi-sra-E-vld-bound-1}).


\ei
To conclude, we only require $m > 1/20$ for the model to satisfy the slow-roll approximation for DBI inflation. For $m>4$ we can conclude that the steeper the potential, the more secure the slow-roll approximation. 

\subs{Validity for particular scalar potentials}
Since we are interested in a few simple models, such as the quadratic potential with $m=2$ and the quartic potential where $m=4$,  we will discuss the validity for these cases.

\ssubs{Quadratic potential}
For the quadratic potential we can write down the range for $\rbF$ and $\rbG$ as
\ba
\rbF(x;m) &=&  2x^2 + \frac{x}{2} + \frac{1}{4}  \in \left(\frac{1}{4}, 1\right) \,,\\
\rbG(x;m) &=& \rbF(x;m) + 2x + \frac{1}{2} \in \left(\frac{3}{4},\frac{5}{2}\right)   \qquad x\in \left(0,\frac{1}{2}\right) \,.
\ea
So we can evaluate $\rm{E}$ according to the Eq.~(\ref{dbi-sra-E-F-G-m-x}), as
\be
|\rm{E}| = \Big|\frac{2\epsilon}{3} \frac{\rbF}{1 - \frac{2\epsilon}{3}\rbG}\Big| < \epsilon \rbF  \label{dbi-sra-E-m2p2}\,,
\ee
due to $\epsilon \ll 1$ during inflation.

\ssubs{Quartic potential}\label{dbi-sra-validity-lp4}

The quartic potential has $m=4$, leading to the results
\ba
\rbF &\equiv&  \frac{1}{4} \,,\\
\rbG &\equiv& \frac{3}{4}  \qquad x\in \left(0,\frac{1}{4}\right) \,.
\ea
So we will have a simple formula for $\rm{E}$ according to the Eq.~(\ref{dbi-sra-E-F-G-m-x}), as
\be
\rm{E} = \frac{\epsilon}{6} \frac{1}{1 - \epsilon/2} \in \left(\frac{\epsilon}{6}, \frac{\epsilon}{3}\right) \label{dbi-sra-E-lp4}\,,
\ee
due to $\epsilon \ll 1$ during inflation.

\subs{Summary}
In our consideration of DBI inflation (\ref{m:DBI}) with polynomial potential form $V(\varphi) = A\varphi^{4-m}$, corresponding to $V(\phi) = A\phi^m$, both Eq.~(\ref{dbi-sra-E-m2p2}) and Eq.~(\ref{dbi-sra-E-lp4}) show that the slow-roll approximation is valid for obtaining results. Also, through the analysis, we have not assumed anything about the sound speed $\cs$; this suggests that even in the limit of $\cs\ll1$, the slow-roll approximation retains its validity in calculating observables such as the power spectrum and its spectral index.

\sect{ $s/2\delta$ for DBI inflation with a Quadratic Potential}\label{dbi-sr-quad-sra}

In this subsection, we present slow-roll calculations for observables in DBI inflation with the quadratic potential.
We start from the slow-roll assumptions
\ba \label{dbi-m2p2-sra}
3H\dot\varphi \simeq \frac{2}{\ls\varphi} (1+y) &\spc& 3H^2 = \frac{1}{\ls\cs\varphi^4}y \,,\\
y = 1 + \alpha \cs \varphi^2 &\spc& \css = 1-\ls{\dot\varphi}^2   \,,\\
\alpha &=& A\ls \,,
\ea
for the quadratic potential in DBI inflation. 

First we present some definitions for a list of small parameters (some of which are derived in terms of $y$ in Section~\ref{sect-dbi-m2p2-sra}), as follows:
\ba
\delta &=& \frac{\dot\varphi}{H\varphi} = \frac{2}{\alpha} \frac{y^2-1}{y} \label{dbi-delta-y} \,,\\
\epsilon &=& -\frac{\dot H}{H^2} = \delta \frac{y+1}{y} = \frac{2}{\alpha} \frac{y^2-1}{y} \frac{y+1}{y} \label{dbi-eps-y}\,,\\
\tilde\eta &=& \frac{\ddot\varphi}{H\dot\varphi} = \delta - \epsilon + \left(\frac{2y^2}{y^2-1} - 1\right) \xi  \label{dbi-teta-y}\,,\\
s &=& \frac{\dot\cs}{H\cs} = -\frac{1-\css}{\css} \frac{\ddot\varphi}{H\dot\varphi} = -\frac{1-\css}{\css}\tilde\eta \label{dbi-s-y} \,,\\
\xi &=& \frac{\dot y}{Hy} = \frac{\dot y}{H(y-1)} \frac{y-1}{y} = (s+2\delta)\frac{y-1}{y} \label{dbi-xi-y} \,,\\
\epsilon &=& \frac{3}{2} \frac{1-\css}{y} \,. \label{dbi-eps-css-y}
\ea
The last equation is the general form of $\epsilon$ for DBI inflation models, independent of the potential. By substituting Eq.~(\ref{dbi-xi-y}) into Eq.~(\ref{dbi-teta-y}) and then into  Eq.~(\ref{dbi-s-y}), the parameters $\tilde\eta$ and $s$ can be written as
\ba
\tilde\eta &=& \delta -\epsilon + \frac{y^2+1}{y(y+1)}(s+2\delta) \label{dbi-teta-y-2} \,,\\
s &=& -\frac{1-\css}{\css} \lrb \delta -\epsilon + \frac{y^2+1}{y(y+1)}(s+2\delta) \rrb \label{dbi-s-y-2}\,.
\ea
We define an auxiliary variable $z=\left|s/2\delta \right|$. According to Eqs.~(\ref{dbi-delta-y}), (\ref{dbi-eps-y}), (\ref{dbi-s-y}), (\ref{dbi-eps-css-y}), and (\ref{dbi-s-y-2}), we can write
\be
z = \left| - \frac{1}{2} \frac{\frac{2y^2-y+1}{y(y+1)}}{\frac{\css}{1-\css} + \frac{y^2+1}{y(y+1)} } \right| = \frac{f(y)}{g(\css) + h(y)} \label{dbi-h-css-y}  \,,
\ee
where we have defined the positive functions
\be
f(y) = \frac{2y^2-y+1}{2y(y+1)} \spc h(y) = \frac{y^2+1}{y(y+1)} \spc g(\css) = \frac{\css}{1-\css} \label{dbi-fgh-m2p2} \,.
\ee
The function $f(y)$ is monotonically increasing as the $y\in (1,\infty)$, and $g(\css)$ is also a monotonically increasing function in $\css \in (0,1]$. Meanwhile $h(y)$ is monotonically decreasing in $y\in(1,1+\sqrt{2})$ then increasing afterwards in $y\in(1+\sqrt{2}, \infty)$. The extremal values for $f(y)$ and $h(y)$ are
\ba
f(y)\Big|_{\rm min} &=& h(y=1) = 0.5 \,,\\
f(y)\Big|_{\rm max} &=& h(y\raw\infty) = 1\,, \\
h(y)\Big|_{\rm min} &=& h(y=\sqrt{2}+1) = 2(\sqrt{2}-1) \sim 0.828 \,,\\
h(y)\Big|_{\rm max} &=& h(y=1) = h(y\raw\infty) = 1\,,
\ea
in the range of $y\in(1,\infty)$.  The only singularity which may occur is located in the term $g(\css)=\css/(1-\css)$. 

We now study two limits of the sound speed $\css$.
\bi
\item[$\star$] $\css \sim 1$\\
In this limit, as $g(\css) \gg 1 \geq h(y)$ we have,
\be
z = \lim_{\css\raw1}\frac{f(y)}{g(\css) + h(y)} \ll 1 \label{dbi-z-css-1} \,.
\ee
We have used that both two functions $f,h$ are bounded in a limited range whatever the value of $y\in(1,\infty)$. Hence we can make the approximation in Eq.~(\ref{sr-dbi-m2p2-dy-deps}) in Section~\ref{sect-dbi-m2p2-sra}. Under this approximation, we conclude that
\be
z \ll 1 \,,
\ee
is always true while $\css \sim O(1)$.

\item[$\star$] $\css \ll 1 $\\
In this limit, $g(\css)$ is much less than one and can even be zero. As $g(\css) < h(y)$ we can approximate $z$ as
\be
z \simeq \frac{f(y)}{h(y)} = \frac{2y^2-y+1}{2(y^2+1)} \label{dbi-z-css-0} \,.
\ee
We find that $z \in (0.5, 1)$ while $y$ ranges from $(1, \infty)$. Therefore in Eq.~(\ref{sr-dbi-m2p2-dy-deps}) in Section~\ref{sect-dbi-m2p2-sra}, this ratio is comparable to the assumed leading term 1. To study the model observables in this case, we cannot use the results which have been obtained in Section~\ref{sect-dbi-m2p2-sra}. Instead we need to include this ratio in Eq.~(\ref{sr-dbi-m2p2-dy-deps}).

\ei

%% file: kmc_dbi_i.bbl
\begin{thebibliography}{40}

\bibitem{LL92} A.~R.~Liddle and D.~H.~Lyth, \emph{COBE, gravitational waves, inflation and extended inflation}, Phys. Lett. B{\bf 291}, 391 (1992) [arXiv:astro-ph/9208007].

\bibitem{Ade:2013uln} 
  P.~A.~R.~Ade {\it et al.}  [Planck Collaboration], \emph{Planck 2013 results. XXII. Constraints on inflation}, arXiv:1303.5082 [astro-ph.CO].

\bibitem{Muk} V.~F.~Mukhanov, \emph{Gravitational instability of the Universe filled with a scalar field}, Pis'ma Zh. Eksp. their. Fiz. {\bf 41}, 402 (1985) [Sov. Phys. JETP Lett. {\bf 41}, 493 (1985)].



\bibitem{Stewart:1993bc} E.~D.~Stewart and D.~H.~Lyth, \emph{A more accurate analytic calculation of the spectrum of cosmological perturbations produced during inflation}, Phys.\ Lett.\ B{\bf 302}, 171 (1993) [gr-qc/9302019].

\bibitem{Gong:2002cx} 
  J.-O.~Gong and E.~D.~Stewart, \emph{The power spectrum for a multicomponent inflaton to second order corrections in the slow roll expansion}, Phys.\ Lett.\ B{\bf 538}, 213 (2002)
  [astro-ph/0202098].

\bibitem{Gong:2001he} 
  J.-O.~Gong and E.~D.~Stewart, \emph{The density perturbation power spectrum to second order corrections in the slow roll expansion}, Phys.\ Lett.\ B{\bf 510}, 1 (2001)
  [astro-ph/0101225].

\bibitem{Choe:2004zg} 
  J.~Choe, J.-O.~Gong, and E.~D.~Stewart, \emph{Second order general slow-roll power spectrum}, JCAP {\bf 0407}, 012 (2004)
  [hep-ph/0405155].

\bibitem{Stewart:2001cd} 
  E.~D.~Stewart, \emph{The spectrum of density perturbations produced during inflation to leading order in a general slow roll approximation}, Phys.\ Rev.\ D {\bf 65}, 103508 (2002)
  [astro-ph/0110322].

\bibitem{Sasaki:1995aw} 
  M.~Sasaki and E.~D.~Stewart, \emph{A general analytic formula for the spectral index of the density perturbations produced during inflation}, Prog.\ Theor.\ Phys.\  {\bf 95}, 71 (1996)
  [astro-ph/9507001].

\bibitem{Huston:2011vt} 
  I.~Huston and K.~A.~Malik, \emph{Second order perturbations during inflation beyond slow-roll}, JCAP {\bf 1110}, 029 (2011)
  [arXiv:1103.0912 [astro-ph.CO]].

\bibitem{Ribeiro:2012ar} 
  R.~H.~Ribeiro, \emph{Inflationary signatures of single-field models beyond slow-roll}, JCAP {\bf 1205}, 037 (2012)
  [arXiv:1202.4453 [astro-ph.CO]].

\bibitem{Adshead:2013zfa} 
  P.~Adshead, W.~Hu, and V.~C.~Miranda, \emph{Bispectrum in single-field inflation beyond slow-roll}, arXiv:1303.7004 [astro-ph.CO].

\bibitem{Burrage:2011hd} 
  C.~Burrage, R.~H.~Ribeiro, and D.~Seery, \emph{Large slow-roll corrections to the bispectrum of noncanonical inflation}, JCAP {\bf 1107}, 032 (2011)
  [arXiv:1103.4126 [astro-ph.CO]].

\bibitem{ArmendarizPicon:1999rj} 
C.~Armendariz-Picon, T.~Damour, and V.~F.~Mukhanov, \emph{K-inflation}, Phys.\ Lett.\  {\bf B458}, 209 (1999)  [hep-th/9904075].

\bibitem{Mizuno:2010ag} 
  S.~Mizuno and K.~Koyama, \emph{Primordial non-Gaussianity from the DBI galileons}, Phys.\ Rev.\ D {\bf 82}, 103518 (2010)
  [arXiv:1009.0677 [hep-th]].

\bibitem{Kobayashi:2010cm} 
  T.~Kobayashi, M.~Yamaguchi, and J.~Yokoyama, \emph{G-inflation: inflation driven by the galileon field}, Phys.\ Rev.\ Lett.\  {\bf 105}, 231302 (2010)
  [arXiv:1008.0603 [hep-th]].

\bibitem{Burrage:2010cu} 
  C.~Burrage, C.~de Rham, D.~Seery, and A.~J.~Tolley, \emph{Galileon inflation}, JCAP {\bf 1101}, 014 (2011)
  [arXiv:1009.2497 [hep-th]].


\bibitem{Li:2012vt} 
  S.~Li and A.~R.~Liddle, \emph{Observational constraints on K-inflation models}, JCAP {\bf 1210}, 011 (2012) 
 [arXiv:1204.6214 [astro-ph.CO]].

\bibitem{Unnikrishnan:2012zu} S. Unnikrishnan, V. Sahni,  and A. Toporensky,  \emph{Refining inflation using non-canonical scalars}, JCAP {\bf 1208}, 018 (2012) [arXiv:1205.0786].

\bibitem{Unnikrishnan2} S. Unnikrishnan and V. Sahni, \emph{Resurrecting power law inflation in the light of Planck results}, JCAP {\bf 1310}, 063 (2013) [arXiv:1305.5260 [astro-ph]].


\bibitem{Garriga:1999vw}
 J.~Garriga and V.~F.~Mukhanov, \emph{Perturbations in k-inflation}, Phys.\ Lett.\  {\bf B458} (1999)  219, [hep-th/9904176].

\bibitem{Kinney} W.~H.~Kinney and K.~Tzirakis, \emph{Quantum modes in DBI inflation: exact solutions and constraints from vacuum selection}, Phys. Rev. D{\bf 77}, 103517 (2008), [arXiv:0712.2043 [astro-ph]].

\bibitem{Ringeval} C.~Ringeval, \emph{Dirac-Born-Infeld and k-inflation: the CMB anisotropies from string theory}. J. Phys. Conf. Ser.  {\bf 203}, 012056 (2010) [arXiv:0910.2167 [astro-ph.CO]].

\bibitem{Hu:2011vr} 
  W.~Hu, \emph{Generalized slow roll for non-canonical kinetic terms}, Phys.\ Rev.\ D {\bf 84}, 027303 (2011)
  [arXiv:1104.4500 [astro-ph.CO]].


\bibitem{Devi:2011qm}  N.~C.~Devi, A.~Nautiyal and A.~A.~Sen, \emph{WMAP constraints on k-inflation}, Phys.\ Rev.\ D {\bf 84}, 103504 (2011)  [arXiv:1107.4911 [astro-ph.CO]].

\bibitem{Mukhanov:2005bu} V.~F.~Mukhanov and A.~Vikman, \emph{Enhancing the tensor-to-scalar ratio in simple inflation}, JCAP {\bf 0602}, 004 (2006) [astro-ph/0512066].

\bibitem{Sen:2002in} 
  A.~Sen, \emph{Tachyon matter}, JHEP {\bf 0207}, 065 (2002)
  [hep-th/0203265].

\bibitem{Sen:2002nu} 
  A.~Sen, \emph{Rolling tachyon}, JHEP {\bf 0204}, 048 (2002)
  [hep-th/0203211].

\bibitem{Gibbons:2002md} 
  G.~W.~Gibbons, \emph{Cosmological evolution of the rolling tachyon}, Phys.\ Lett.\ B{\bf 537}, 1 (2002)
  [hep-th/0204008].

\bibitem{Kofman:2002rh} 
  L.~Kofman and A.~D.~Linde, \emph{Problems with tachyon inflation}, JHEP {\bf 0207}, 004 (2002)
  [hep-th/0205121].

\bibitem{Fairbairn:2002yp} 
  M.~Fairbairn and M.~H.~G.~Tytgat, \emph{Inflation from a tachyon fluid?}, Phys.\ Lett.\ B{\bf 546}, 1 (2002)
  [hep-th/0204070].

\bibitem{Piao:2002vf} 
 Y.~-S.~Piao, R.~-G.~Cai, X.~-m.~Zhang and Y.~-Z.~Zhang, \emph{Assisted tachyonic inflation},
 Phys.\ Rev.\ D {\bf 66}, 121301 (2002)
 [hep-ph/0207143].


\bibitem{Alishahiha:2004eh}  M. Alishahiha, E. Silverstein, and D. Tong, \emph{DBI in the sky}, Phys.\ Rev.\ D{\bf 70},  123505 (2004)  [hep-th/0404084].

\bibitem{Silverstein:2003hf} 
  E.~Silverstein and D.~Tong, \emph{Scalar speed limits and cosmology: Acceleration from D-cceleration}, Phys.\ Rev.\ D {\bf 70}, 103505 (2004)
  [hep-th/0310221].

\bibitem{Peiris:2007gz} H.~V.~Peiris, D.~Baumann, B.~Friedman, and A. Cooray, \emph{Phenomenology of D-Brane Inflation with General Speed of Sound}, Phys.\ Rev.\ D {\bf 76}, 103517 (2007) [arXiv:0706.1240 [astro-ph]].

\bibitem{Lorenz:2008et} L.~Lorenz, J.~Martin, and C.~Ringeval, \emph{K-inflationary Power Spectra in the Uniform Approximation}, Phys.\ Rev.\ D {\bf 78}, 083513 (2008) [arXiv:0807.3037 [astro-ph]].

\bibitem{MRV} J.~Martin, C.~Ringeval and V.~Vennin, \emph{K-inflationary Power Spectra at Second Order}, JCAP {\bf 1306}, 021 (2013) [arXiv:1303.2120 [astro-ph]].

\bibitem{Powell:2008bi} B.~A.~Powell, K.~Tzirakis, and W. H. Kinney, \emph{Tensors, non-Gaussianities, and the future of potential reconstruction}, JCAP {\bf 0904}, 019 (2009) [arXiv:0812.1797 [astro-ph]].

\bibitem{Bean:2007eh} 
  R.~Bean, X.~Chen, H.~Peiris, and J.~Xu, \emph{Comparing infrared Dirac-Born-Infeld brane inflation to observations}, Phys.\ Rev.\ D {\bf 77}, 023527 (2008)
  [arXiv:0710.1812 [hep-th]].

\bibitem{Ade:2013ng} 
  P.~A.~R.~Ade {\it et al.}  [Planck Collaboration], \emph{Planck 2013 results. XXIV. Constraints on primordial non-Gaussianity}, arXiv:1303.5084 [astro-ph.CO].

\bibitem{lambert-wiki}
 \emph{Lambert W function}, \url{https://en.wikipedia.org/wiki/Lambert_W_function}

\bibitem{lambertw-HH}
 A.~Hoorfar and M.~Hassanij, \emph{Inequalities on the Lambert W function and hyperpower function}, Inequal. Pure and Appl. Math., 9:2, Art.\ 51 (2008).

\bibitem{Liddle:2003as} 
  A.~R.~Liddle and S.~M.~Leach, \emph{How long before the end of inflation were observable perturbations produced?}, Phys.\ Rev.\ D {\bf 68}, 103503 (2003)
  [astro-ph/0305263].

\bibitem{Lewis:2002ah}  A.~Lewis and S.~Bridle, \emph{Cosmological parameters from CMB and other data: A Monte Carlo approach}, Phys.\ Rev.\ D {\bf 66}, 103511 (2002),  [astro-ph/0205436]. Code at \url{http://cosmologist.info/cosmomc/}

\end{thebibliography}
